\documentclass[11pt]{article}
\pdfoutput=1 
\usepackage{amssymb}
\usepackage{amsmath}
\usepackage{amstext}
\usepackage{graphicx,epsfig}
\usepackage{epsfig}
\usepackage{verbatim} 
\usepackage{fancybox}
\usepackage{color}
\usepackage{ulem}
\usepackage{enumitem}
\usepackage{youngtab}
\usepackage{subfigure}
\usepackage{bbm}
\usepackage{parskip}
\usepackage[numbers,sort&compress]{natbib}
\usepackage{ytableau}
\usepackage[utf8]{inputenc}
\usepackage{tikz}
\usepackage{simplewick}
\usetikzlibrary{positioning, trees,decorations, decorations.markings, decorations.pathmorphing, arrows, shapes.geometric, snakes}

\newcommand{\Comment}[1]{{}}
\definecolor{darkblue}{rgb}{0.15,0.35,0.55}
\definecolor{reddish}{rgb}{0.65, 0.2, 0.2}
\usepackage[linktocpage=true]{hyperref}
\hypersetup{
colorlinks=true,
citecolor=darkblue,
linkcolor=reddish,
urlcolor=darkblue,
pdfauthor={},
pdftitle={},
pdfsubject={}
}

\newcommand{\be}{\begin{equation}}
\newcommand{\ee}{\end{equation}}
\newcommand{\nn}{\nonumber}
\newcommand{\ccdot}{\! \cdot \!}
\newcommand{\intN}{\int_{\mathcal{N}}\ }

\newcommand{\bea}{\begin{eqnarray}}
\newcommand{\eea}{\end{eqnarray}}

\numberwithin{equation}{section}

\begin{document}
\tikzset{
    photon/.style={decorate, decoration={snake}, draw=magenta},
vector/.style={decorate, decoration={snake}, draw=blue},
    graviton/.style={decorate, decoration={snake}, draw=black},
 sgal/.style={decorate , dashed, draw=black},
 scalar/.style={decorate , draw=black},
  mgraviton/.style={decorate, draw=black,
    decoration={coil,amplitude=4.5pt, segment length=7pt}}
    electron/.style={draw=blue, postaction={decorate},
        decoration={markings,mark=at position .55 with {\arrow[draw=blue]{>}}}},
    gluon/.style={decorate, draw=magenta,
        decoration={coil,amplitude=4pt, segment length=5pt}} 
}

\renewcommand{\thefootnote}{\fnsymbol{footnote}}
~
\vspace{1.75truecm}
\thispagestyle{empty}
\begin{center}
{\LARGE \bf{
Unitarization from Geometry
}}
\end{center} 

\vspace{1cm}
\centerline{\Large James Bonifacio,${}^{\rm a,}$\footnote{\href{mailto:james.bonifacio@case.edu}{\texttt{james.bonifacio@case.edu}}}
and Kurt Hinterbichler,${}^{\rm a,}$\footnote{\href{mailto:kurt.hinterbichler@case.edu} {\texttt{kurt.hinterbichler@case.edu}}}
}
\vspace{.5cm}

\centerline{{\it ${}^{\rm a}$CERCA, Department of Physics,}}
 \centerline{{\it Case Western Reserve University, 10900 Euclid Ave, Cleveland, OH 44106}} 
 \vspace{.25cm}

\vspace{1cm}
\begin{abstract}
\noindent
We study the perturbative unitarity of scattering amplitudes in general dimensional reductions of Yang--Mills theories and general relativity on closed internal manifolds. For the tree amplitudes of the dimensionally reduced theory to have the expected high-energy behavior of the higher-dimensional theory, the masses and cubic couplings of the Kaluza--Klein states must satisfy certain sum rules that ensure there are nontrivial cancellations between Feynman diagrams. These sum rules give constraints on the spectra and triple overlap integrals of eigenfunctions of Laplacian operators on the internal manifold and can be proven directly using Hodge and eigenfunction decompositions. One consequence of these constraints is that there is an upper bound on the ratio of consecutive eigenvalues of the scalar Laplacian on closed Ricci-flat manifolds with special holonomy. This gives a sharp bound on the allowed gaps between Kaluza--Klein excitations of the graviton that also applies to Calabi--Yau compactifications of string theory.

\end{abstract}

\newpage

\setcounter{tocdepth}{3}
\tableofcontents
\renewcommand*{\thefootnote}{\arabic{footnote}}
\setcounter{footnote}{0}

\section{Introduction}

Since the work of Kaluza and Klein \cite{Kaluza:1921tu,Klein:1926tv}, the idea of compact extra dimensions has been pervasive in physics.  A simple higher-dimensional theory containing only a few degrees of freedom when  placed on a product spacetime with a compact internal manifold becomes a complicated tower of infinitely many degrees of freedom when looked at from the point of view of the lower-dimensional spacetime.  The masses of the particles in the towers are determined by the eigenvalues of various Laplacian operators on the internal space and their couplings are determined by various multiple overlap integrals of the eigenfunctions.

If the higher-dimensional theory is General Relativity (GR), i.e., an interacting theory of a massless spin-2 particle, then the lower-dimensional Kaluza--Klein (KK) theory will contain a tower of massive spin-2 modes whose masses are determined by the eigenvalues of the scalar Laplacian on the internal manifold.  These massive spin-2 modes will be fully interacting, with their interactions determined by higher-dimensional GR.  
It is difficult to directly construct an interacting theory of a massive spin-2 particle whose tree amplitudes are well-behaved at high energies.  This difficulty shows up in the scattering of the longitudinal modes, which do not decouple as in the massless theory and typically have the worst high-energy behavior.  
For example, naively adding a mass term to GR leads to a theory in which the tree amplitude for scattering the longitudinal modes of four gravitons grows with the center-of-mass energy $E$ as $\sim E^{10}$, thus violating perturbative unitarity at a relatively low scale \cite{ArkaniHamed:2002sp}.  This can be improved to $\sim E^{6}$ by carefully choosing the interaction terms \cite{ArkaniHamed:2002sp,Creminelli:2005qk,deRham:2010ik} to be those of the de Rham--Gabadadze--Tolley (dRGT) theory \cite{deRham:2010kj} (see \cite{Hinterbichler:2011tt,deRham:2014zqa} for reviews), and this is the best that can be done with a single massive graviton \cite{Schwartz:2003vj, Bonifacio:2018vzv}, even when including a massless graviton or any finite number of additional vectors and scalars \cite{Bonifacio:2018aon, Bonifacio:2019mgk}.

The massive gravitons that arise from KK compactifications are an exception to these constraints.  In GR in any dimension, the four-point graviton amplitude behaves  at high energies as $\sim E^2$.  Since the dimensionally reduced KK theory with all of the massive modes kept is just a rewriting of the higher-dimensional theory, the lower-dimensional amplitudes should also behave as $\sim E^2$ at high energies where the infrared effects of the compactification are negligible.  Thus the infinite tower of massive gravitons and other modes, with the specific interactions dictated by the KK theory, should somehow evade the bounds present for a single massive graviton.
In this paper, we will study the interactions among the particles in the KK tower and calculate their amplitudes to see exactly how the infinite tower with the specific KK couplings achieves the improved high-energy behavior.  We will be able to see how it works for a completely general compactification on any closed smooth manifold.\footnote{The case of compactifications on a one-dimensional manifold has been considered recently in Refs.~\cite{Chivukula:2019rij, Chivukula:2019zkt}.}

\begin{figure}[h!]
\begin{center}
\epsfig{file=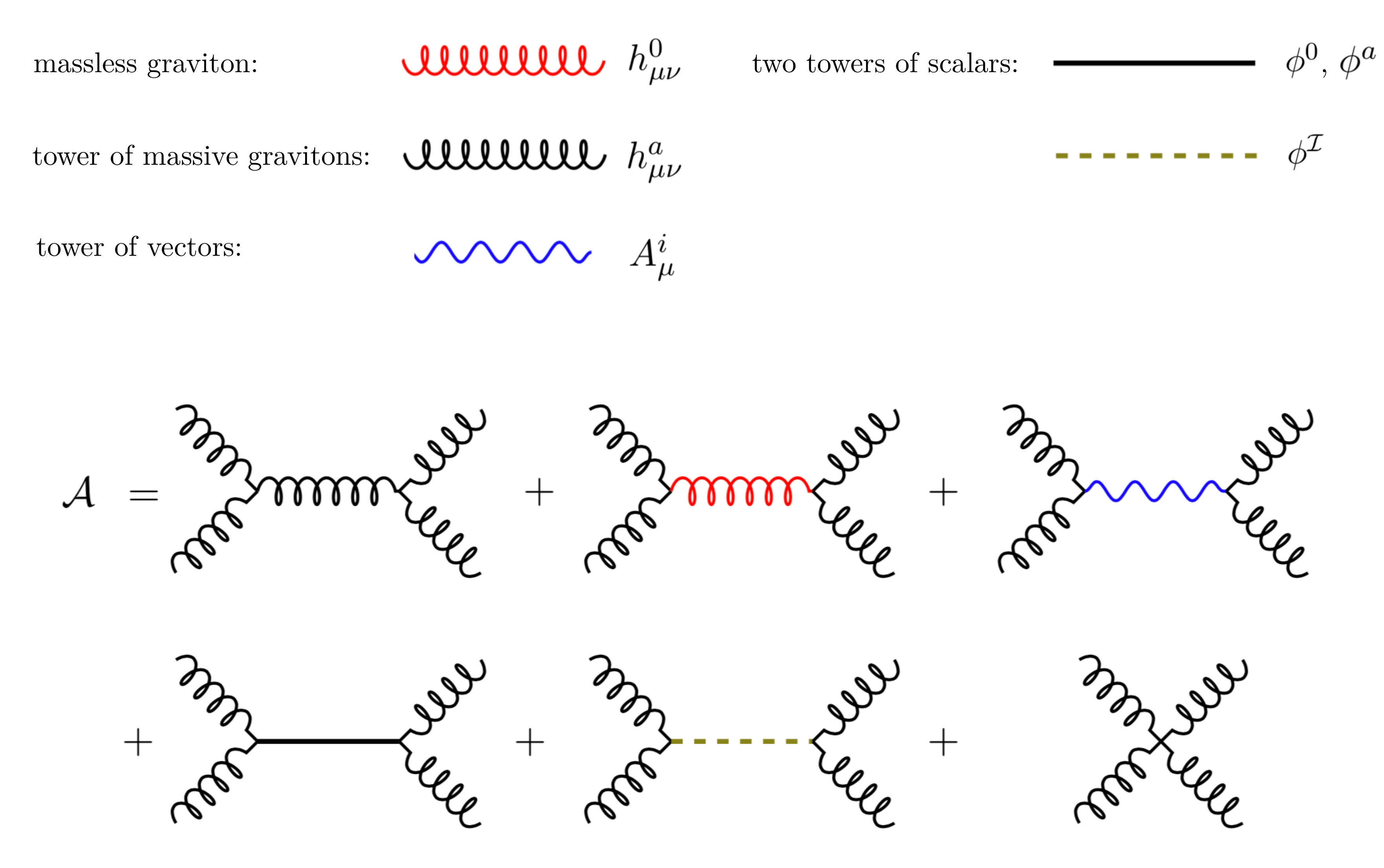,height=3.5in,width=5.0in}
\caption{The spectrum of GR KK reduced on a general closed Ricci-flat manifold and a schematic representation of the four-point tree-level massive graviton amplitudes.}
\label{figure1}
\end{center}
\end{figure}

In more detail, when gravity is compactified on an internal manifold ${\cal N}$, the spectrum of the resulting theory on the lower-dimensional spacetime ${\cal M}$ contains a single massless spin-2 graviton, corresponding to the zero mode of the scalar Laplacian, a tower of massive spin-2 particles with masses given by the non-zero eigenvalues of the scalar Laplacian, a tower of vectors corresponding to the transverse eigenmodes of the vector Laplacian on ${\cal N}$, and two towers of massive scalars, corresponding to eigenmodes of the scalar and Lichnerowicz Laplacians on ${\cal N}$. In Figure \ref{figure1} we show this spectrum together with a schematic representation of the tree amplitudes of four massive gravitons in the KK theory.

The amplitudes for the longitudinal modes have a high-energy expansion which starts at $E^{10}$,
\be  {\cal A}=\alpha_{10} E^{10}+\alpha_8 E^8+\alpha_6 E^6+\alpha_4 E^4+\alpha_2 E^2+\dots ,\ee
since this is the generic scaling of a massive spin-2 amplitude.
The coefficients $\alpha_i$ are combinations of the various kinematic factors and couplings that enter the Feynman diagrams.
In the KK theory the coefficients $\alpha_{10}$, $\alpha_{8}$, $\alpha_{6}$, and $\alpha_{4}$ must all somehow vanish, since the high-energy behavior should be improved to $\sim E^2$ to match the higher-dimensional theory.
As we will see, the canceling of these coefficients and the resulting softening of the high-energy behavior of the KK theory occurs due to various nontrivial sum rules that must be satisfied by the masses and couplings.  These masses and couplings are in turn determined by geometric data of the internal manifold consisting of the eigenvalues and overlap integrals of eigenfunctions of various Laplacians on the manifold.

There is an analogy between this geometric data and the data describing a Conformal Field Theory (CFT) that we will find useful to keep in mind.
Consider the collection of real eigenmodes $\mathcal{O}_I$ of the various Laplacians defined on $\mathcal{N}$, which satisfy eigenvalue equations of the form
\be
\Delta \mathcal{O}_I = \lambda_I \mathcal{O}_I.
\ee
In addition to the eigenfunctions of the scalar Laplacian, this also includes the eigenmodes of Laplacians acting on higher-rank tensor fields, such as the Hodge Laplacian on transverse $p$-forms and the Lichnerowicz Laplacian on transverse traceless symmetric tensors. 
The analogy is  that the eigenmodes $\mathcal{O}_I$ are like CFT operators and their eigenvalues $\lambda_I$ are like the conformal scaling dimensions of these operators. 

Taking the analogy further, we can think of integrals over $\mathcal{N}$ of products of eigenfunctions as the analogues of CFT correlation functions. For example, the eigenfunctions can always be chosen to be orthonormal, 
\be
\intN \mathcal{O}_{I_1}  \mathcal{O}_{I_2} = \delta_{I_1 I_2},\label{orthonormeq1}
\ee
which is analogous to the statement that CFT two-point functions can be diagonalized. 
We can similarly form ``three-point functions" as triple overlap integrals
\be
 \intN \mathcal{O}_{I_1}  \mathcal{O}_{I_2}  \mathcal{O}_{I_3} \equiv  g_{I_1 I_2 I_3},\label{trheepointdefe}
\ee
and higher $k$-point functions
\be \intN \mathcal{O}_{I_1}  \mathcal{O}_{I_2} \cdots \mathcal{O}_{I_k} \equiv g_{I_1 I_2\cdots I_k}.\label{higherpointdefee}\ee
We can also consider integrals involving covariant derivatives acting on the eigenmodes, which are analogous to correlators of descendant operators in the CFT.

The eigenfunctions we consider are complete, i.e. they form an orthonormal basis for the associated spaces of functions on ${\cal N}$, so we can write normalizable fields as linear combinations of them.  In particular, in the cases we consider we can expand the product $\mathcal{O}_{I_1}  \mathcal{O}_{I_2} $ of two eigenmodes as a sum over eigenmodes, with orthonormality implying that the coefficients in the expansion are three-point coefficients \eqref{trheepointdefe},
\be
 \mathcal{O}_{I_1}  \mathcal{O}_{I_2}=\sum_{I_3} g_{I_1 I_2}^{\ \  \ \ I_3}\mathcal{O}_{I_3} .\label{OPEexpansione}
\ee
This is analogous to the operator product expansion (OPE) of a CFT.  The higher-point functions \eqref{higherpointdefee} with $k\geq4$ can all be reduced to sums of products of three-point coefficients by repeatedly using \eqref{OPEexpansione},
so the eigenvalues $\lambda_I$ and cubic couplings $g_{I_1 I_2 I_3}$ constitute some of the basic geometric data of a manifold.\footnote{The conformal dimensions and OPE coefficients completely determine the CFT, so it is natural to ask whether a smooth manifold is determined by the geometric data of eigenvalues and cubic couplings of, say, the Hodge Laplacians.  It is known, for example, that the eigenvalues of the $p$-form Laplacians do not by themselves uniquely determine a manifold, since there exist manifolds that are  $p$-isospectral for all $p$~\cite{Milnor542}.  It would be interesting to know if including the triple overlap integrals does provide sufficient information to uniquely determine a manifold.}
For example, the quartic coefficients $g_{I_1 I_2 I_3 I_4}$ can be written as 
\be 
g_{I_1 I_2 I_3 I_4} \equiv \int_{\mathcal{N}}
 \contraction{}{\mathcal{O}_{I_1} }{}{\mathcal{O}_{I_2} }
\contraction{\mathcal{O}_{I_1} \mathcal{O}_{I_2} }{\mathcal{O}_{I_3} }{}{\mathcal{O}_{I_4} }
\mathcal{O}_{I_1} \mathcal{O}_{I_2} \mathcal{O}_{I_3} \mathcal{O}_{I_4} 
= \sum_I g_{I_1 I_2}{}{}^{ I} g_{I_3 I_4 I}  , \label{ffirstwickintrore}
\ee
where we use the Wick contraction notation to indicate that we expand the indicated pair of fields using \eqref{OPEexpansione}.
We could also have expanded in the other channels,
\be \label{eq:OPEintegral}
\int_{\mathcal{N}} \contraction{}{\mathcal{O}_{I_1}}{}{\mathcal{O}_{I_2}}
\contraction{\mathcal{O}_{I_1}\mathcal{O}_{I_2}}{\mathcal{O}_{I_3}}{}{\mathcal{O}_{I_4}}
\mathcal{O}_{I_1} \mathcal{O}_{I_2} \mathcal{O}_{I_3} \mathcal{O}_{I_4}
=
\int_{\mathcal{N}} \contraction{}{\mathcal{O}_{I_1}}{\mathcal{O}_{I_2}}{\mathcal{O}_{I_3}}
\contraction[2ex]{\mathcal{O}_{I_1}}{\mathcal{O}_{I_2}}{\mathcal{O}_{I_3}}{\mathcal{O}_{I_4}}
\mathcal{O}_{I_1} \mathcal{O}_{I_2} \mathcal{O}_{I_3} \mathcal{O}_{I_4}
=
\int_{\mathcal{N}} \contraction[2ex]{}{\mathcal{O}_{I_1}}{\mathcal{O}_{I_2}\mathcal{O}_{I_3}}{\mathcal{O}_{I_4}}
\contraction{\mathcal{O}_{I_1}}{\mathcal{O}_{I_2}}{}{\mathcal{O}_{I_3}}
\mathcal{O}_{I_1} \mathcal{O}_{I_2} \mathcal{O}_{I_3} \mathcal{O}_{I_4}.
\ee
There are thus multiples ways to write the quartic coupling in terms of cubic couplings, which give associativity relation constraints on the cubic couplings,
\be \label{eq:assocO}
g_{I_1 I_2 I_3 I_4} = \sum_I g_{I_1 I_2}{}{}^{ I} g_{I_3 I_4 I} = \sum_I g_{I_1 I_3 }{}{}^{ I} g_{I_2 I_4 I}   = \sum_I g_{I_1 I_4}{}{}^{ I} g_{I_2 I_3 I}.
\ee 
These are analogous to the associativity relations in a CFT, upon which the conformal bootstrap is built (see \cite{Poland:2018epd} for a recent review).  

In the KK theory, the $k$\textsuperscript{th} overlap integrals determine the $k$-point couplings among the various modes in the KK towers.
As we will see, the various associativity relations satisfied by these overlap integrals, and hence by the couplings, are equivalent to the unitarity sum rules and so they are the mechanism behind the improved high-energy behavior of the tree amplitudes in the KK theory. 
Since we have the freedom to compactify gravity on any Ricci-flat manifold, the sum rules must correspond to universally true, purely mathematical statements about closed Ricci-flat manifolds, and hence they can imply various nontrivial mathematical facts about such manifolds.  

For example, we will find that the sum rules imply that the ratios of consecutive nonzero eigenvalues of the scalar Laplacian on a closed Ricci-flat manifold with nonnegative Lichnerowicz eigenvalues are bounded above by four,
\be \label{eq:eigenbound1}
\frac{\lambda_{k+1}}{\lambda_k} \leq 4,
\ee
where $\lambda_k$ is the $k$\textsuperscript{th} nonzero eigenvalue.
This bound  is new as far as we know and applies to all known closed Ricci-flat manifolds. It implies that the gaps between the massive spin-2 excitations of the graviton cannot be large relative to their masses, a result that also applies to the low-energy limits of smooth Calabi--Yau compactifications of string theory and $G_2$ compactifications of M-theory.   If there were such gaps, we could integrate out of all the KK modes above some fixed scale and obtain an effective field theory with a finite number of massive modes and a strong coupling scale that is parametrically larger than their masses. However, Eq.~\eqref{eq:eigenbound1} implies that we cannot obtain effective theories of bi-gravity \cite{Hassan:2011zd} or multi-gravity \cite{Hinterbichler:2012cn} in this way.\footnote{However, they can be obtained from discretized extra dimensions \cite{deRham:2013awa}.}  

We emphasize that both the higher- and lower-dimensional theories are effective field theories that become strongly coupled at some ultraviolet (UV) energy scale.
An apparent puzzle is how the cutoff scales of the theories match in different dimensions.\footnote{The matching of counterterms in higher- and lower-dimensional theories is considered in Refs.~\cite{Duff:1982gj, Duff:1982wm}.} For example, the lower-dimensional massive spin-2 amplitudes behave as well as GR amplitudes and so they seem to become strongly coupled around the lower-dimensional Planck scale, rather than the expected higher-dimensional Planck scale. The resolution is to consider amplitudes involving states that are normalized superpositions of all states appearing below the cutoff, as in Refs.~\cite{SekharChivukula:2001hz, Chivukula:2019rij}. Such amplitudes have the same energy growth but are parametrically larger than amplitudes for single-particle states, since there are parametrically many particles below the cutoff, and indeed become strong at the higher-dimensional Planck scale.

\begin{figure}[h!]
\begin{center}
\epsfig{file=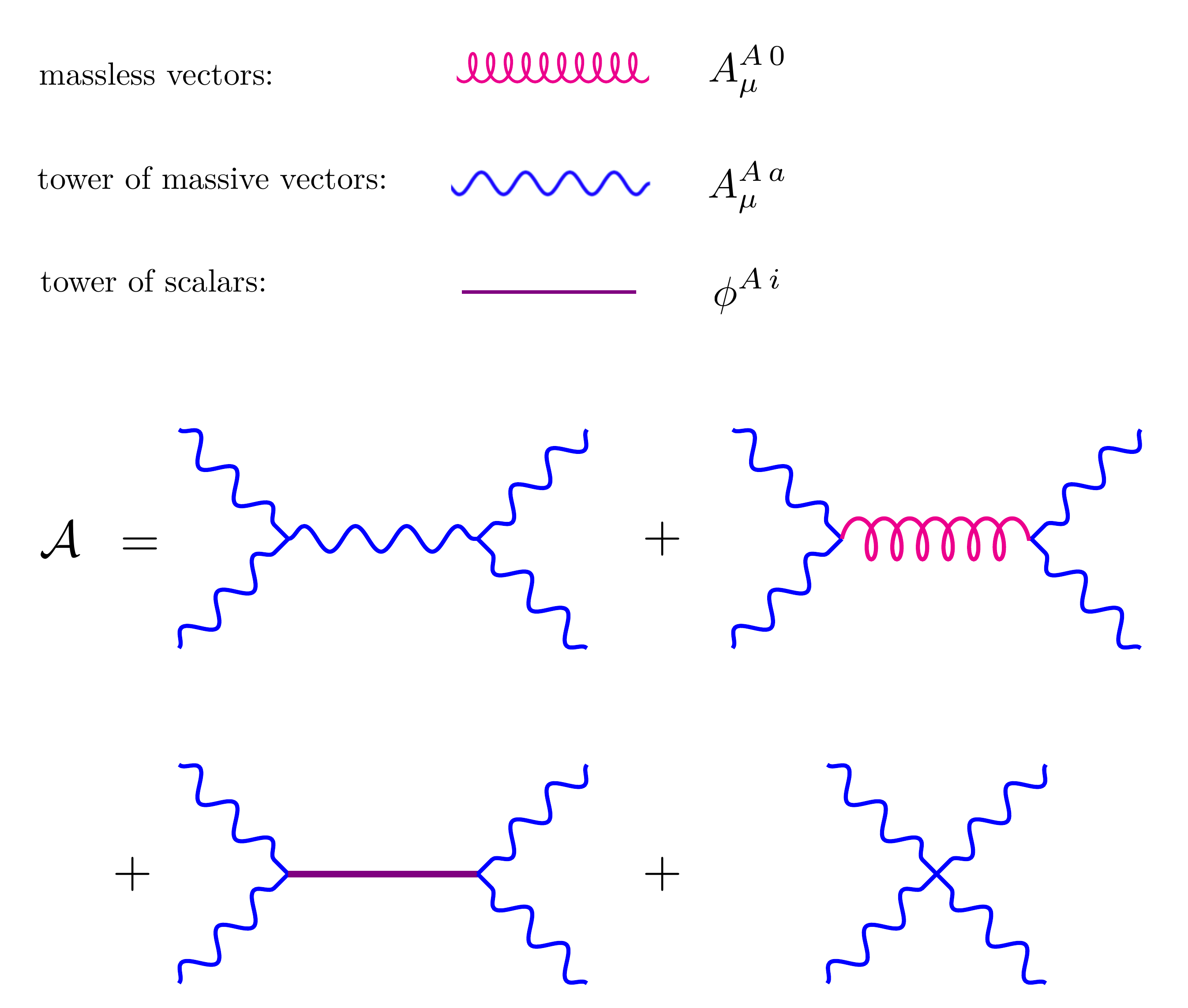,height=4.0in,width=4.5in}
\caption{The spectrum of a YM theory KK reduced on a general manifold and a schematic representation of the four-point tree-level massive vector amplitudes.}
\label{figure2}
\end{center}
\end{figure}

A similar story goes through if we start with a higher-dimensional non-abelian Yang--Mills (YM) theory.
Dimensionally reducing gives a lower-dimensional YM theory coupled to an infinite tower of massive adjoint vectors plus a tower of adjoint scalars.
Generically, the four-point tree amplitudes of longitudinally polarized massive vectors grow like $\sim E^4$ at high energies,
\be  {\cal A}=\alpha_4 E^4+\alpha_2 E^2+\alpha_0 E^2+\cdots .\ee
However, to match the high-energy behavior of higher-dimensional YM theory, the massive vectors in the lower-dimensional KK theory must have tree amplitudes that are bounded at high energies. This implies the existence of sum rules that cause the coefficients $\alpha_4$ and $\alpha_2$ to vanish. For YM theories there is no restriction on the internal manifold, unlike for gravity where the internal manifold has to be an Einstein space, so these sum rules must apply to any geometry.
The spectrum of the KK theory and the massive vector amplitudes that must realize these cancellation are shown schematically in Figure \ref{figure2}.

The rest of the paper is structured as follows: in Section~\ref{sec:YM} we study scattering amplitudes of dimensionally reduced YM theory and in Section~\ref{sec:GR} we study scattering amplitudes of dimensionally reduced GR. In each case we start with the higher-dimensional theory and perform a Hodge plus eigenfunction decomposition of the fields, which allows us to work with general internal manifolds. After dealing with the gauge freedom, we find the relevant lower-dimensional interactions by integrating over the internal manifold. The resulting vertices are then used to compute various four-point amplitudes, from which we deduce sum rules by imposing that the high-energy growth matches that of the higher-dimensional theory. Some general consequences of these sum rules are then discussed before we show how to derive them directly from geometry. We also discuss some explicit examples of manifolds to demonstrate explicitly their consistency with the sum rules and their consequences. We conclude and summarize our results in Section \ref{sec:conclusion}.  Some known results about the eigenvalues of the scalar Laplacian are reviewed in Appendix~\ref{app:bounds}.

\vspace{.5cm}
\noindent {\bf Conventions:} We use the mostly plus signature convention, $(-, +, +, \ldots)$. The lower-dimensional spacetime $\mathcal{M}$ has dimension $d>2$, indices $\mu, \nu,\dots$, and coordinates $x^{\mu}$; the internal space $\mathcal{N}$ has dimension $N$, indices $n , m,\dots$, and coordinates $y^{n}$; and the full spacetime is their product $\mathcal{M} \times \mathcal{N}$ which has indices $M, N$ and coordinates $X^M$. We use the Einstein summation convention for the various spacetime indices and for YM color indices $A_1, A_2, \ldots$, but not for the indices $a_1, a_2, \ldots$, $i_1, i_2, \ldots$, and $\mathcal{I}_1, \mathcal{I}_2, \ldots$, which label states in the KK tower.

\section{Yang--Mills theory}
\label{sec:YM}
In this section we consider the perturbative unitarity of scattering amplitudes in a general dimensional reduction of pure YM theory on a closed manifold down to a lower-dimensional flat spacetime. The case of a single compact extra dimension has been studied previously in Refs.~\cite{Hill:2000mu, Dicus:2000hm, SekharChivukula:2001hz, Csaki:2003dt, Abe:2003vg, SekharChivukula:2008mj}.

\subsection{Higher-dimensional theory}

We start with a pure YM theory of a compact gauge group on a $D$-dimensional background spacetime that is a product of $d$-dimensional Minkowski spacetime $\mathcal{M}$ and an $N$-dimensional Riemannian manifold $\mathcal{N}$.  We take ${\cal N}$ to be smooth and compact without boundary, i.e. closed and with no defects or branes.  We also assume that $\mathcal{N}$ is connected and orientable. The full metric is
\be
ds^2 = {G}_{A B} dX^A dX^B= \eta_{\mu \nu} dx^{\mu} dx^{\nu} + \gamma_{mn} dy^m dy^n \, ,
\ee
where $x^{\mu}$ are coordinates on $\mathcal{M}$, $y^m$ are coordinates on $\mathcal{N}$, $\eta_{\mu\nu}$ is the flat metric on $\mathcal{M}$, and $\gamma_{mn}$ is the general curved metric on $\mathcal{N}$. 

The YM Lagrangian is
\be \label{eq:LYM}
\mathcal{L} = -\frac{1}{4}\sqrt{-G}  F_{M N}^{A_1} F^{M N}_{A_1},
\ee
where $A_1$ is the color index and the field strength is
\be
F_{M N}^{A_1} = \partial_M V^{A_1}_N - \partial_N V^{A_1}_M +g_D f^{A_1}{}_{ A_2 A_3} V_M^{A_2} V_N^{A_3}.
\ee
Here $g_D$ is the $D$-dimensional gauge coupling, which has mass dimension $[g_D]=-{D-4\over 2}$, and $f^{}{}_{ A_1A_2 A_3}$ are the completely anti-symmetric structure constants of the gauge group. Color indices are raised and lowered with $\delta_{A_1 A_2}$.
The action is invariant under the gauge transformations
\be
\delta V_M^{A_1} = \partial_M \Lambda^{A_1} + g_D f^{A_1}{}_{ A_2 A_3} V_{M}^{A_2} \Lambda^{A_3},
\ee
where $\Lambda^{A_j}$ are the scalar gauge parameters. 

Expanding out \eqref{eq:LYM} gives the quadratic, cubic, and quartic terms:
\begin{align}
\mathcal{L}_{(2)}  & =-\frac{1}{4} \sqrt{-G} \left(\partial_M V^{A_1}_N - \partial_N V^{A_1}_M \right)^2 ,\label{lagymhd2}\\
\mathcal{L}_{(3)} & =-\frac{1}{2}\sqrt{-G} \,g_D f^{A_1}{}_{ A_2 A_3} \left(\partial^M V_{A_1}^N - \partial^N V_{A_1}^M \right)  V_{M}^{A_2} V_{N}^{A_3},\label{lagymhd3}\\
\mathcal{L}_{(4)} & =-\frac{1}{4} \sqrt{-G} \,g_D^2 f^{A}{}_{ A_1 A_2} f_{A A_3 A_4} V^{A_1} \ccdot V^{A_3} V^{A_2}\ccdot V^{A_4}, \label{lagymhd4}
\end{align}
where $V \ccdot V \equiv V_M V^M$.

\subsubsection{Hodge and eigenfunction decompositions}
\label{sec:YMHodge}
To integrate out the extra dimensions, we first perform a Hodge and eigenfunction decomposition of the fields. This is the generalization of the familiar Fourier decomposition used for a single compact extra dimension and allows us to make the reduction on a general internal manifold. The decomposition is given by 
\be \label{eq:hodge}
V^A_{M}(x,y) =
\left( \begin{matrix}
\sum_a A^{Aa}_{\mu}(x) \psi_a(y) +\frac{1}{\sqrt{V}}A^{A0}_{\mu}(y)\\
 \sum_a A^{A a}(x) \partial_n \psi_a(y)+\sum_i \phi^{A i}(x) Y_{n, i}(y) 
\end{matrix} \right).
\ee
The different terms in this decomposition are as follows.
The functions $\psi_a$ are real orthonormal eigenmodes of the scalar Laplace--Beltrami operator $\Delta$ on $\mathcal{N}$ that have positive eigenvalues $\lambda_a>0$ labeled by the discrete index $a$ (or $a_1$, $a_2, \dots$),
\be
\Delta  \psi_a \equiv -\Box \psi_a  = \lambda_a \psi_a, \quad \intN \psi_{a_1} \psi_{a_2} = \delta_{a_1 a_2},
\ee
where $\intN$ denotes the integral over $\mathcal{N}$ with the standard volume form $\sqrt{\gamma} d^N y$. The zero mode is given by the normalized constant $V^{-1/2}$,  where $V$ is the volume of ${\cal N}$, and is separated out explicitly in Eq.~\eqref{eq:hodge} and thus is not included in the index $a$.
The $\psi_a$ and the zero mode together form a basis for the vector space of square-integrable functions on $\mathcal{N}$. 

The vectors
$Y_{n, i}$ are the real and orthonormal transverse eigenmodes of the vector Hodge Laplacian on $\mathcal{N}$ with eigenvalues $\lambda_i\geq0$ labeled by the discrete index $i$ (or $i_1$, $i_2, \dots$),
\be
\Delta Y_{n, i} \equiv - \Box Y_{n, i} + R_n{}^m Y_{m, i}= \lambda_i Y_{n, i}, \quad \nabla^n Y_{n, i} =0, \quad\intN Y^n_{i_1} Y_{n, i_2} = \delta_{i_1 i_2},
\ee
where $R_{mn}$ is the Ricci curvature of ${\cal N}$.
These form a basis of normalizable transverse vectors on $\mathcal{N}$. The vectors with zero eigenvalue correspond to harmonic vectors, which are included in the index $i$. The number of independent harmonic vectors is given by the first Betti number of $\mathcal{N}$, $b_1(\mathcal{N})$.

The coefficients of the eigenfunctions appearing in Eq.~\eqref{eq:hodge} are functions of $x$ that will correspond to $d$-dimensional fields after dimensionally reducing.  We discuss them below.

\subsubsection{Gauge fixing}
We can similarly decompose the higher-dimensional gauge parameters,
\be
\Lambda^A(x,y) =\sum_a \Lambda^{Aa}(x) \psi_a(y) +\frac{1}{\sqrt{V}}\Lambda^{A0}(x).
\ee
At leading order in the fields, the $d$-dimensional fields $A^{Aa}$ transform under this gauge transformation as
\be
\delta A^{Aa} =  \Lambda^{A a}+\cdots.
\ee
We can thus partially fix the gauge symmetry by using the gauge parameters $\Lambda^{A a}$ to fix the gauge 
\be A^{Aa}=0.\label{YMgaugechoicee}\ee
The $\Lambda^{Aa}$ are determined iteratively in an expansion in powers of the fields from this gauge choice.\footnote{An alternative to algebraically  fixing the St\"uckelberg symmetry is to form gauge-invariant combinations order-by-order in powers of the fields, e.g., as in  Refs.~\cite{Skenderis:2006uy, Caldarelli:2018azk}. The lowest-order gauge-invariant combination in this case is $\tilde{A}^{Aa}_{\mu}  \equiv A^{Aa}_{\mu} -\partial_{\mu} A^{A a}$.  This gets complicated at higher orders, but the net result is equivalent to eliminating $A^{Aa}$.} 

Once this gauge is fixed, the leftover gauge symmetry is the zero mode of the gauge symmetry, parameterized by $\Lambda^{A 0}$. This acts on the vector zero modes as a lower-dimensional YM gauge symmetry,
\be
 \delta A_{\mu}^{A_1 0}  =  \partial_{\mu} \Lambda^{A_1 0}+\frac{g_D}{\sqrt{V}} f^{A_1}{}_{ A_2 A_3} A_{\mu}^{A_2 0} \Lambda^{A_3 0}.
\ee
From this we read off that the effective $d$-dimensional YM coupling $g_d$ is given by 
\be
g_d = \frac{ g_D}{\sqrt{V}}.\label{YMcouplingsfirstde}
\ee
The remaining fields transform linearly under the adjoint representation of the zero mode gauge symmetry,
\begin{align}
\delta A^{A_1 a}_{\mu} & =g_d f^{A_1}{}_{A_2 A_3} A_{\mu}^{A_2a} \Lambda^{A_30}, \\
\delta \phi^{A_1 i} & = g_d f^{A_1}{}_{A_2 A_3} \phi^{A_2 i} \Lambda^{A_3 0}.
\end{align}
From the point of view of the lower-dimensional YM symmetry, these fields are all matter fields in the adjoint representation. 

\subsection{Lower-dimensional interactions}
We can now substitute the decomposition~\eqref{eq:hodge} into the higher-dimensional action and integrate over the internal manifold using the orthonormality of the eigenmodes to find the lower-dimensional Lagrangian.

\subsubsection{Spectrum}
Performing this procedure on the quadratic part of the higher-dimensional Lagrangian \eqref{lagymhd2} and using our gauge choice \eqref{YMgaugechoicee}, we obtain the quadratic lower-dimensional Lagrangian \cite{Hinterbichler:2013kwa},
\begin{align}
\mathcal{L}_{(2)} & = -\frac{1}{4}\left(\partial_{\mu} A^{A0}_{\nu}-\partial_{\nu} A^{A 0}_{\mu}\right)^2 -\frac{1}{2}\sum_a \left( \frac{1}{2}\left(\partial_{\mu} A^{A a}_{\nu}-\partial_{\nu} A^{A a}_{\mu}\right)^2+ \lambda_a \left(A_{\mu}^{Aa} \right)^2 \right)\nonumber \\
& -\frac{1}{2}\sum_i \left(\left(\partial_{\mu} \phi^{A i} \right)^2+ \lambda_i \left(\phi^{A i}\right)^2 \right).
\end{align}
From this we can read off the following degrees of freedom:
\begin{enumerate}
\item A single multiplet of massless vectors, $A_{\mu}^{A0}$.
\item  A tower of massive vector multiplets $A_{\mu}^{A a}$ with squared masses $m^2_a = \lambda_a$, one for every non-constant eigenmode of the scalar Laplacian on $\mathcal{N}$.
\item A tower of scalar multiplets $\phi^{A i}$ with squared masses $m_i^2=\lambda_i$, one for each transverse vector eigenmode of the vector Laplacian of $\mathcal{N}$, including a massless scalar multiplet for each harmonic vector.
\end{enumerate}

\subsubsection{Cubic interactions\label{firstcubicsubsec}}
We can similarly extract the cubic interactions from \eqref{lagymhd3} after dimensionally reducing. For the four-point massive vector amplitude we are primarily interested in, we need cubic interactions with at least two massive vectors, which enter the exchange diagrams as shown in Figure \ref{figure2}. These interactions are
\begin{align}
\mathcal{L}_{AAA} & =- g_D f_{A_1 A_2 A_3} \sum_{a_1, a_2, a_3}  g_{a_1 a_2 a_3}\partial_{[\mu} A_{\nu]}^{A_1 a_1} A^{\mu A_2 a_2} A^{ \nu A_3 a_3} , \label{eq:LAAA} \\
\mathcal{L}_{AAA^0} & =-g_d f_{A_1 A_2 A_3} \sum_{a_1} \left(\partial_{[\mu} A_{\nu]}^{A_1 a_1} \left( A^{\mu A_2}_{a_1} A^{\nu A_3 0}+  A^{\mu A_2 0} A^{\nu A_3}_{ a_1}\right)+ \partial_{[\mu} A_{\nu]}^{ A_1 0} A^{\mu A_2}_{ a_1} A^{\nu A_3 a_1} \right) , \label{glueglueve}\\
\mathcal{L}_{AA \phi} & =g_D f_{A_1A_2 A_3} \sum_{a_1, a_2, i_3}g_{a_1 a_2 i_3}A^{A_1 a_1} \ccdot A^{A_2 a_2} \phi^{A_3 i_3} , 
\end{align}
where $A \ccdot A \equiv A_{\mu} A^{\mu}$ and we have defined two sets of cubic couplings in terms of triple overlap integrals of the eigenmodes,
\begin{align}
g_{a_1 a_2 a_3} & \equiv \int_{\mathcal{N}} \psi_{a_1} \psi_{a_2} \psi_{a_3}, \label{eq:tripleint1}\\
g_{a_1 a_2 i_3} &\equiv \int_{\mathcal{N}}  \partial^{n_1} \psi_{a_1} \psi_{a_2} Y_{n_1, i_3} \label{eq:tripleint2}.
\end{align}
The cubic coupling $g_{a_1 a_2 a_3}$ is fully symmetric in its indices, whereas $g_{a_1 a_2 i_3}$ is anti-symmetric in its first two indices.
Since the gluons $A_\mu^{A0}$ correspond to zero modes, their cubic interactions with two massive vectors $A_{\mu}^{A_1 a_1}$ and $A_{\mu}^{A_2 a_2}$ are proportional to $\delta_{a_1 a_2}$, so there is no additional triple overlap integral for this interaction.  

To calculate scattering amplitudes we need the various cubic vertices obtained from these interactions.
The full off-shell vertices used in the Feynman rules with all momenta incoming are given by
\begin{align}
\mathcal{V}(1_A^{A_1 a_1}, \, 2_A^{A_2 a_2}, \, 3_A^{A_3 a_3}) & = g_D f^{A_1 A_2 A_3} g_{a_1a_2a_3}\left( \epsilon_1 \ccdot \epsilon_2 \, \epsilon_3 \ccdot (p_1-p_2)+\rm{cyclic}\right), \\
\mathcal{V}(1_A^{A_1 a_1}, \, 2_A^{A_2 a_2}, \, 3_{A}^{A_3 0})& = g_d  f^{A_1 A_2 A_3} \delta_{a_1 a_2} \left( \epsilon_1 \ccdot \epsilon_2 \, \epsilon_3 \ccdot (p_1-p_2)+\rm{cyclic}\right), \\
\mathcal{V}(1_A^{A_1 a_1}, \, 2_A^{A_2 a_2}, \, 3_{\phi}^{A_3 i_3})&  =2 i g_D f^{A_1 A_2 A_3} g_{a_1 a_2 i_3} \, \epsilon_1 \ccdot \epsilon_2 ,\label{eq:AAphi}
\end{align}
where ``cyclic" denotes the $(123)$ and $(132)$ cyclic permutations of the first terms.

\subsubsection{Quartic interaction}

To compute the contact diagram shown in Figure \ref{figure2} we also need the quartic interaction of four KK vectors. This can be obtained from \eqref{lagymhd4} and is given by
\begin{align}
\mathcal{L}_{AAAA}  =&- \frac{ g_D^2}{4}  f^{A_1 A_2}{}{}_A f^{ A_3 A_4 A} \sum_{a_1, a_2, a_3, a_4} g_{a_1 a_2 a_3 a_4} A^{A_1 a_1} \ccdot A^{A_3 a_2} A^{A_2 a_3}\ccdot A^{A_4 a_4} , \label{eq:LAAAA} 
\end{align}
where we have defined the quartic coupling
\be \label{eq:lambda1234}
g_{a_1 a_2 a_3 a_4} \equiv \intN \psi_{a_1} \psi_{a_2} \psi_{a_3} \psi_{a_4},
\ee
which is fully symmetric in its indices.

The corresponding vertex is given by
\begin{align}
& \mathcal{V}(1_A^{A_1 a_1}, \, 2_A^{A_2 a_1}, \, 3_A^{A_3a_3},\, 4_A^{A_4 a_4})  = - i g_D^2  g_{a_1 a_2 a_3 a_4}  \Big[ \epsilon_1 \ccdot \epsilon_2 \, \epsilon_3 \ccdot \epsilon_4  \left(f^{A_1 A_3}{}{}_A f^{A_2 A_4 A}+f^{A_1 A_4}{}{}_A f^{A_2 A_3 A} \right)\nn \\
&  +\epsilon_1 \ccdot \epsilon_3 \, \epsilon_2 \ccdot \epsilon_4 \left(f^{ A_1 A_2}{}{}_A f^{A_3 A_4 A}-f^{A_1 A_4}{}{}_A f^{A_2 A_3 A} \right)  -\epsilon_1 \ccdot \epsilon_4 \, \epsilon_2 \ccdot \epsilon_3  \left(f^{A_1 A_2}{}{}_A f^{A_3 A_4 A}+f^{A_1 A_3}{}{}_A f^{A_2 A_4 A} \right)  \Big].
\end{align}
This can be further simplified using the Jacobi identity but we leave it in this form to manifest the particle exchange symmetries.

\subsection{Expansion of eigenfunction products}
\label{sec:completeness}

We present here some useful expansions and identities that we will need later.  We first define two more triple overlap integrals that will be needed later, 
\begin{align} \label{eq:giii}
g_{i_1 i_2 i_3} & \equiv \int_{\mathcal{N}} \left(\partial_{n} Y_{m, i_1}-\partial_{m} Y_{n, i_1} \right) Y^n_{i_2} Y^m_{i_3},  \\
\label{eq:gaii}
g_{a_1 i_2 i_3} & \equiv \int_{\mathcal{N}} \psi_{a_1} Y^{n}_{i_2} Y_{n, i_3},
\end{align}
which are antisymmetric and symmetric in their last two indices, respectively.
Using the Hodge decomposition plus the fact that eigenfunctions form a basis, we can always expand the product of two eigenfunctions as a sum over eigenfunctions with coefficients involving triple overlap integrals. For example, using scalar completeness we get the following useful relations: 
\begin{align} 
\psi_{a_1} \psi_{a_2} & = \sum_a g_{a_1 a_2}{}{}^{a} \psi_{a}+\frac{1}{V} \delta_{a_1 a_2},\label{eq:completeness1} \\
\partial^n \psi_{a_1} \partial_n \psi_{a_2} & = \sum_a \frac{\left(\lambda_{a_1} + \lambda_{a_2} - \lambda_a \right)}{2} g_{a_1 a_2}{}{}^{a} \psi_a + \frac{\lambda_{a_1}}{V} \delta_{a_1 a_2}, \label{eq:completeness6} \\
Y^n_{i_1}Y_{n, i_2} & = \sum_a g^a{}_{ i_1 i_2} \psi_{a}+\frac{1}{V} \delta_{i_1 i_2} , \label{eq:completeness2} \\
\partial^n \psi_{a_1} Y_{n, i_2} & = -\sum_a g^a{}_{ a_1 i_2} \psi_a. \label{eq:completeness4}
\end{align}

Similarly, using the vector Hodge decomposition and the completeness of eigenfunctions we get
\begin{align}
\psi_{a_1}Y_{n, i_2} & = \sum_i g_{a_1 i_2}{}{}^i Y_{n, i} + \sum_a \frac{g^a{}_{ a_1 i_2}}{\lambda_{a}} \partial_n \psi_{a}, \label{eq:completeness3} \\
\partial_n \psi_{a_1} \psi_{a_2}  & = \sum_i g_{a_1 a_2}{}{}^i Y_{n, i} + \sum_a \frac{\left( \lambda_{a_1} - \lambda_{a_2}+\lambda_a\right) }{2 \lambda_a}g_{a_1 a_2}{}{}^a \partial_n \psi_a . \label{eq:completeness5}
\end{align}

The eigenvalues and triple overlap integrals of eigenfunctions, such as $(\lambda_{a_1}, g_{a_1 a_2 a_3})$, constitute some of the basic geometric data of a compact manifold.
As discussed in the introduction, we can make an analogy to CFT data: the eigenfunctions are like conformal primary operators of a CFT, the eigenvalues are like the conformal dimensions, and the triple overlap integrals are like the OPE coefficients. 
The above decompositions of products of eigenfunctions are then analogous to OPE expansions.

We can use these expansions to reduce all higher-point integrals to products of eigenvalues and triple overlap integrals, just as the OPE in a CFT can be used to reduce all higher-point correlators. For example, the quartic couplings $g_{a_1 a_2 a_3 a_4}$ defined in Eq.~\eqref{eq:lambda1234}  can be written as 
\be \label{eq:quartic1234}
g_{a_1 a_2 a_3 a_4} \equiv \int_{\mathcal{N}}
 \contraction{}{\psi_{a_1}}{}{\psi_{a_2}}
\contraction{\psi_{a_1}\psi_{a_2}}{\psi_{a_3}}{}{\psi_{a_4}}
\psi_{a_1} \psi_{a_2} \psi_{a_3} \psi_{a_4}
= \sum_a g_{a_1 a_2}{}{}^{ a} g_{a_3 a_4 a} +\frac{1}{V}  \delta_{a_1 a_2} \delta_{a_3 a_4} ,
\ee
where we use the Wick contraction notation to indicate that we expand the indicated pair of fields using the above eigenvalue expansion. We could also have expanded in the other channels,
\be \label{eq:OPEintegral}
\int_{\mathcal{N}} \contraction{}{\psi_{a_1}}{}{\psi_{a_2}}
\contraction{\psi_{a_1}\psi_{a_2}}{\psi_{a_3}}{}{\psi_{a_4}}
\psi_{a_1} \psi_{a_2} \psi_{a_3} \psi_{a_4}
=
\int_{\mathcal{N}} \contraction{}{\psi_{a_1}}{\psi_{a_2}}{\psi_{a_3}}
\contraction[2ex]{\psi_{a_1}}{\psi_{a_2}}{\psi_{a_3}}{\psi_{a_4}}
\psi_{a_1} \psi_{a_2} \psi_{a_3} \psi_{a_4}
=
\int_{\mathcal{N}} \contraction[2ex]{}{\psi_{a_1}}{\psi_{a_2}\psi_{a_3}}{\psi_{a_4}}
\contraction{\psi_{a_1}}{\psi_{a_2}}{}{\psi_{a_3}}
\psi_{a_1} \psi_{a_2} \psi_{a_3} \psi_{a_4},
\ee
so there are two ways to write the quartic coupling in terms of cubic couplings in addition to Eq.~\eqref{eq:quartic1234},
\be \label{eq:assoc}
g_{a_1 a_2 a_3 a_4} = \sum_a g_{a_1 a_3 }{}{}^{ a} g_{a_2 a_4 a} +\frac{1}{V}  \delta_{a_1 a_3} \delta_{a_2 a_4}   = \sum_a g_{a_1 a_4}{}{}^{ a} g_{a_2 a_3 a} + \frac{1}{V}   \delta_{a_1 a_4} \delta_{a_2 a_3}.
\ee

Invoking the CFT analogy again, the relations \eqref{eq:OPEintegral} are the geometric analogues of the scalar bootstrap equations derived from OPE associativity.  We can similarly find relations involving spinning eigenfunctions by expanding other quartic integrals in multiple ways. For example, from the associativity relation
\be
\int_{\mathcal{N}}
\contraction{}{ \psi_{a_1}}{}{ \psi_{a_2}}
\contraction{ \psi_{a_1} \psi_{a_2}}{Y_{n, i_3}}{}{Y_{n, i_4} }
 \psi_{a_1} \psi_{a_2} Y_{n, i_3} Y_{n, i_4}
 =
\int_{\mathcal{N}}
\contraction{}{ \psi_{a_1}}{ \psi_{a_2}}{Y_{n, i_3}}
\contraction[2ex]{ \psi_{a_1}}{ \psi_{a_2}}{Y_{n, i_3}}{Y_{n, i_4} }
 \psi_{a_1} \psi_{a_2} Y_{n, i_3} Y_{n, i_4} 
\ee
we get the identity
\be
\sum g_{a_1 a_2}{}{}^{ a} g_{a i_3 i_4} + \frac{1}{V}  \delta_{a_1 a_2} \delta_{i_3 i_4 } 
= \sum_i g_{a_1 i_3}{}{}^{ i} g_{ a_2 i_4 i} +\sum_a \lambda_{a}^{-1} g^a{}_{ a_1 i_3} g_{a a_2 i_4} .  \label{eq:assoc2}
\ee
These associativity relations and their generalizations will end up being equivalent to the sum rules we derive later from perturbative unitarity constraints.

\subsection{Amplitudes and sum rules}
We now calculate some four-point amplitudes.
We take the $d$-dimensional momenta to be
\be
p_j^{\mu}(p_j) = (E_j, p_j \sin \theta_j, 0, \dots, 0, p_j \cos \theta_j ),
\ee
where $j\in (1,2,3,4)$ labels the particle, $p_j$ is the magnitude of the 3-momentum of the $j$\textsuperscript{th} particle and $\theta_j$ is its scattering angle, 
\be
\theta_1 =0, \quad \theta_2=\pi, \quad \theta_3=\theta, \quad \theta_4 =\theta- \pi.
\ee
The energy $E_j$ satisfies $E_j^2=p_j^2+m_j^2$ where $m_j$ is the mass of the $j$\textsuperscript{th} particle.
The longitudinal vector polarizations are
\be \label{eq:spin1pol}
\epsilon^{\mu}_L (p_j) = \frac{1}{m_j}(p_j, E_j \sin \theta_j, 0, \dots, 0, E_j \cos \theta_j ).
\ee

\subsubsection{Simple four-point vector scattering}
We consider first the following inelastic process, where the incoming particles share a color and KK flavor index, as do the outgoing particles:
\be \label{eq:A1A2A1A2}
A_L^{A_1 a_1} A_L^{A_1 a_1} \rightarrow A_L^{A_3 a_3} A_L^{A_3 a_3}.
\ee
The diagrams that contribute to this amplitude are the $t$- and $u$-channel exchange of massive KK vectors, scalars, and gluons, plus the contact diagram.

For generic masses and coupling constants with the interactions derived above, the high-energy behavior of this amplitude is $\sim E^4$. However, we expect that the high-energy behavior should be the same as for YM in higher dimensions, namely $\sim E^0$. This implies that the couplings and masses should satisfy certain sum rules that lead to this improved high-energy behavior. 

Indeed, the leading $E^4$ part of the amplitude manifestly vanishes after we use the associativity relations in Eq.~\eqref{eq:OPEintegral} to evaluate the quartic coupling, so these relations give our first examples of sum rules. From now on we will automatically write the quartic coupling in terms of the cubic couplings using Eq.~\eqref{eq:quartic1234}.

Setting to zero the subleading order $E^2$ part of the amplitude gives a more complicated sum rule,
\be \label{eq:sumrule1}
4 \sum_i g_{a_1 a_3 i}^2+ \sum_a \lambda_{a}^{-1} \left((\lambda_{a_1}-\lambda_{a_3})^2+2(\lambda_{a_1}+\lambda_{a_3})\lambda_{a}-3 \lambda_{a}^2 \right)g_{a_1 a_3 a}^2+4 V^{-1}\lambda_{a_1} \delta_{a_1 a_3}=0,
\ee
where we have removed the overall color and coupling constant factors. The first term comes from the contributions of the exchanged scalars, the second term comes from the exchanged vectors with masses $m_a^2= \lambda_a$, and the last term is due to gluon exchange. 
If we further set  $a_1=a_3$, so the external particles all have the same KK index, then the contribution from scalar exchange drops out and the sum rule becomes
\be \label{eq:sumrule1b}
\sum_a \left(4\lambda_{a_1}-3 \lambda_{a} \right)g_{a_1 a_1 a}^2+4V^{-1}\lambda_{a_1} =0.
\ee
This has exactly the same form as the sum rule that was obtained in Ref.~\cite{Csaki:2003dt} by considering the unitarity of amplitudes in YM compactified on an interval with general boundary conditions.
Below we will show how to directly prove these sum rules for any closed manifold, which confirms that the tree amplitudes describing the process \eqref{eq:A1A2A1A2} are indeed bounded at high energy. 

These sum rules have consequences for the spectrum and couplings of the KK theory.
From Eq.~\eqref{eq:sumrule1b} we can easily deduce something interesting. Since the second term, coming from gluon exchange, is strictly positive, the sum over $a$ must contain a negative term, which is only possible if a sufficiently heavy vector is exchanged.  In terms of geometric quantities, the requirement is that for every real scalar eigenfunction $\psi_ {a_1}$ there must exist some $a^*$ such that 
\be \label{eq:YMbound}
g_{a_1 a_1 a^*} \equiv \intN \psi_{a_1}^2 \psi_{a^*} \neq 0 \quad {\rm and } \quad  \frac{4}{3} \lambda_{a_1} < \lambda_{a^*} .
\ee
This bound was pointed out already in Ref.~\cite{Duff:1989ea} when discussing the lack of a consistent finite massive truncation of the KK tower of the graviton.

In terms of particle interactions, Eq.~\eqref{eq:YMbound} says that each KK vector with mass $m_{a_1}$ must have a nonvanishing cubic interaction with a heavier vector of mass $m_{a^*}$ such that
\be
  \frac{2}{\sqrt{3}} m_{a_1} < m_{a^*}.
\ee
By applying the same argument to this heavier particle, we can see that to unitarize its four-point amplitudes we need an even heavier particle, which then necessitates the existence of an even heavier particle, and so on, as in the five-dimensional case~\cite{SekharChivukula:2001hz, Csaki:2003dt}.  This implies that no finite number of massive modes can decouple from the infinite tower without worsening the high-energy behavior.

There is a similar bound that follows from the more general sum rule in Eq.~\eqref{eq:sumrule1}. Namely, for each $a_1$ and $a_2$ such that there exists an $i^*$ for which $g_{a_1 a_2 i^*} \neq0$, there must exist some $a^*$ such that
\be
 g_{a_1 a_2 a^*} \neq 0 \quad {\rm and } \quad\frac{1}{3} \left( \lambda_{a_1} + \lambda_{a_2} + 2 \sqrt{\lambda_{a_1}^2-\lambda_{a_1} \lambda_{a_2} + \lambda_{a_2}^2} \right) <  \lambda_{a^*} .
\ee
In terms of particle interactions, this implies that any two massive KK vectors that couple to a scalar must also couple to a third vector that is heavier than either of them.

\subsubsection{General four-point vector scattering}
We now consider the general four-point process with longitudinally-polarized external vectors,
\be \label{eq:A1A2A3A4}
A_L^{A_1 a_1} A_L^{A_2 a_2} \rightarrow A_L^{A_3 a_3} A_L^{A_4 a_4},
\ee
where now all of the color indices are distinct. The diagrams contributing to this process are shown schematically in Figure~\ref{figure2}.

The leading $E^4$ part again vanishes after using the associativity relations \eqref{eq:OPEintegral}. At order $E^2$ there are four new sum rules, which generalize Eq.~\eqref{eq:sumrule1}. To find these sum rules we independently set to zero the angle-independent part of the amplitude and the part proportional to $\cos \theta$. We also use the Jacobi identity,
\be \label{eq:jacobi}
 f^{A_1 A_2}{}{}_{A} f^{A_3 A_4 A} - f^{A_1 A_3}{}{}_{A} f^{A_2 A_4 A}+ f^{A_1 A_4}{}{}_{A} f^{A_2 A_3 A} =0,
\ee
to reduce to two independent color factors.
We can write the resulting four sum rules as follows:
\begin{subequations} \label{eq:YMsumrules}
\begin{align}
 &\sum_a \left[\left(\lambda  -\lambda_a\right) \left( g_{a_1 a_3}{}{}^{a} g_{a_2 a_4 a}+g_{a_1 a_4 }{}{}^{a} g_{a_2 a_3 a}\right)-\left(\lambda +\lambda_a\right)g_{a_1 a_2}{}{}^{a} g_{a_3 a_4 a}\right]  \nn \\
&+ \lambda  V^{-1}  \left( \delta_{a_1 a_3}\delta_{a_2 a_4}+\delta_{a_1 a_4}\delta_{a_2 a_3}-\delta_{a_1 a_2}\delta_{a_3 a_4}\right) =0, \\
&\sum_a  \lambda_a^{-1}\left( (\lambda_{a_1}-\lambda_{a_3})(\lambda_{a_2}-\lambda_{a_4})+\lambda \lambda_a-3\lambda_a^2 \right)  g_{a_1 a_3}{}{}^{a} g_{a_2 a_4 a} + 4 \sum_i g_{a_1 a_3}{}{}^{i} g_{a_2 a_4 i} \nn \\
& +\lambda V^{-1} \delta_{a_1 a_3}\delta_{a_2 a_4} + \left( a_3 \leftrightarrow a_4 \right) =0, \label{eq:4ptrule2} \\
 & \sum_a  \lambda_a^{-1}  \left(  (\lambda_{a_1}-\lambda_{a_3})(\lambda_{a_2}-\lambda_{a_4})-3\lambda \lambda_a+\lambda_a^2 \right)  g_{a_1 a_3}{}{}^{a} g_{a_2 a_4 a}+ 4 \sum_i  g_{a_1 a_3}{}{}^{i} g_{a_2 a_4 i} \nn \\
& -3 \lambda  V^{-1}\delta_{a_1 a_3}\delta_{a_2 a_4} - \left( a_3 \leftrightarrow a_4 \right) =0, \\
& \sum_a \lambda_a^{-1} \left((\lambda_{a_1}-\lambda_{a_3})(\lambda_{a_2}-\lambda_{a_4})-2\lambda_a^2\right)g_{a_1 a_3}{}{}^{a} g_{a_2 a_4 a}+ 4\sum_i g_{a_1 a_3}{}{}^{i} g_{a_2 a_4 i}  - \left( a_3 \leftrightarrow a_4 \right)   \nn \\
 &  -3\sum_a \lambda_a^{-1} (\lambda_{a_1}-\lambda_{a_2})(\lambda_{a_3}-\lambda_{a_4})g_{a_1 a_2}{}{}^{a} g_{a_3 a_4 a} - 12\sum_i g_{a_1 a_2}{}{}^{i} g_{a_3 a_4 i} =0,
\end{align}
\end{subequations}
where we have defined
\be
\lambda \equiv \lambda_{a_1}+ \lambda_{a_2}+ \lambda_{a_3}+ \lambda_{a_4}.
\ee
These are the general sum rules for the four-point scattering of longitudinally-polarized vectors.\footnote{We have checked in $d=4$ that we get no additional constraints by scattering vectors with transverse polarizations. This is expected since transverse polarizations lead to softer high-energy behavior, so the perturbative unitarity constraint is easier to satisfy.}

After enforcing these sum rules, the leading nonvanishing part of the amplitude is independent of energy and can be written as
\begin{align}
& \mathcal{A}(1_{A_L}^{A_1 a_1}, \, 2_{A_L}^{A_2 a_2}, \, 3_{A_L}^{A_3 a_3}, \, 4_{A_L}^{A_4 a_4})   =  \frac{ 2 g_D^2}{m_1 m_2 m_3 m_4}\left(2 f^{A_1 A_3}{}{}_A f^{A_2 A_4 A} + (\cos \theta -1)  f^{A_1 A_2}{}{}_A f^{A_3 A_4 A} \right) \times \nn \\ 
&  \intN\left( \frac{ \partial \psi_{a_1} \ccdot \partial \psi_{a_4} \partial \psi_{a_2} \ccdot \partial \psi_{a_3} }{1+ \cos \theta} + \frac{ \partial \psi_{a_1} \ccdot \partial \psi_{a_3} \partial \psi_{a_2} \ccdot \partial \psi_{a_4}}{1-\cos \theta}-\frac{1}{2} \partial \psi_{a_1} \ccdot \partial \psi_{a_2} \partial \psi_{a_3} \ccdot \partial \psi_{a_4} \right)+\mathcal{O}(E^{-2}), \label{eq:YME2}
\end{align}
where $\theta$ is the scattering angle and we have simplified using the Jacobi identity \eqref{eq:jacobi}. This agrees with the result of Ref.~\cite{SekharChivukula:2001hz} for the particular case of an orbifold compactification on $S^1/\mathbb{Z}_2$, up to an overall factor due to differing normalizations. 

\subsubsection{Proving the sum rules}

We now show how the remaining sum rules can be proven using the eigenvalue and Hodge decompositions to evaluate two-derivative integral identities. As a simple example, we can derive the sum rule in Eq.~\eqref{eq:sumrule1} from the associativity relation
\be
\int_{\mathcal{N}}
\contraction{}{\psi_{a_1}}{}{\psi_{a_3}}
\contraction{\psi_{a_1} \psi_{a_3}}{\partial_{m} \psi_{a_1}}{}{\partial^{m} \psi_{a_3}}
\psi_{a_1} \psi_{a_3} \partial_{m} \psi_{a_1} \partial^m \psi_{a_3}
=
\int_{\mathcal{N}}
\contraction[2ex]{}{\psi_{a_1}}{\psi_{a_3} \partial_{m} \psi_{a_1}}{ \partial^{m} \psi_{a_3}}
\contraction{\psi_{a_1}}{ \psi_{a_3}}{}{\partial_{m} \psi_{a_1}}
\psi_{a_1} \psi_{a_3} \partial_{m} \psi_{a_1} \partial^m \psi_{a_3} \, .
\ee
The more general sum rules can similarly be derived by evaluating integrals of total derivative combinations of eigenfunctions. In fact, in turns out that they all correspond to different ways of evaluating the simple identity
\be
\int_{\mathcal{N}} \nabla^m\left(\partial_m \psi_{a_1} \psi_{a_2} \psi_{a_3} \psi_{a_4}\right) =0.
\ee
A systematic method to verify this is to use the relations in Section~\ref{sec:completeness} to convert the sum rules \eqref{eq:YMsumrules} into integrals over $\mathcal{N}$ and then check that the variations of the integrands vanish. 

\subsubsection{Strong coupling scale}

A YM theory in $D$ dimensions has a coupling $g_D$ with mass dimension $-{D-4\over 2}$.  For $D>4$ this coupling has a negative mass dimension and so the theory is non-renormalizable, i.e. it is an effective theory with a strong coupling scale $\Lambda$ set by the coupling,
\be \label{eq:gDstrong}
\Lambda \sim g_D^{-{2\over D-4}}.
\ee
We can see from Eq.~\eqref{eq:YME2} that the amplitudes in the dimensionally reduced theory go to a constant set by the lower-dimensional YM coupling $g_d$,\footnote{This assumes that various dimensionless combinations of geometrical quantities are not parametrically small, e.g., 
ratios of eigenvalues or couplings. 
For this schematic argument we assume that such terms are of order one.}
\be
\mathcal{A}(E) \sim g_d^2\, .
\ee 
Thus the theory naively seems to become strongly coupled around  the scale $\sim g_d^{-{2\over d-4}}$.
 However, this cannot be correct since the dimensionally reduced theory is equivalent to the higher-dimensional theory and the higher-dimensional theory becomes strong at the much lower scale $\Lambda$ given in Eq.~\eqref{eq:gDstrong}. The reason this scale is lower is because of the relation $g_d=V^{-1/2}g_D$ and the fact that the volume of the internal manifold should satisfy ${ V^{-1/N}}\ll g_D^{-{2\over D-4}}$ for the geometry to be trustable within the higher-dimensional theory.

The way to see the lower cutoff from the lower-dimensional point of view is by considering the scattering of states that are normalized superpositions of KK modes, as in Refs.~\cite{SekharChivukula:2001hz, Chivukula:2019rij}.  The four-point scattering of superposed states of $n$ modes gives an additional factor of $n$ in the amplitude,
\be 
{\cal A}^{\rm superposition}(E) \sim n g_d^{2} \label{speccuteYM}.
\ee
 To maximize the amplitude we should make $n$ as large as possible.  From Weyl's law \eqref{eq:Weyl} on the asymptotic distribution of eigenvalues on the $N$-dimensional internal manifold, the number of spin-1 modes below the cutoff $\Lambda$ is given at leading order by
\be 
n(\Lambda) \sim V \Lambda^N + \mathcal{O}(\Lambda^{N-1})\,,\label{nrelaeqn}
\ee  
so this is the maximum number of spin-1 states that we can scatter.
Since the compactification is only trustworthy when $V^{-1/N} \ll \Lambda$, this number is necessarily much larger than one.
We can determine the cutoff $\Lambda$ from the condition that the amplitude saturates the $d$-dimensional unitarity bound ${\cal A}^{\rm superposition}(E) \sim 1/E^{d-4}$ when $E=\Lambda$ and $n = n(\Lambda)$.  
This gives
\be
{\cal A}^{\rm superposition}(\Lambda) \sim {1\over \Lambda^{d-4}}  \implies \Lambda \sim g_D^{-{2\over D-4}},
\ee
confirming that the strong coupling scale matches the higher-dimensional theory.
The strong coupling scale is thus parametrically lower due to the large number of states in the KK tower. 

\subsection{Other four-point processes}
\label{sec:other4pt}
The four-point amplitude with external longitudinally-polarized massive vectors has the most divergent high-energy growth.  Other amplitudes also have growing parts that must cancel, so it is also interesting to consider four-point processes with different external particles to see if these give any additional constraints involving the other masses and couplings. Altogether there are 15 different processes to consider, up to crossing.

To compute these amplitudes we need three additional nonzero cubic interactions beyond those of Section \ref{firstcubicsubsec},
\begin{align}
\mathcal{L}_{\phi \phi \phi}  =& -\frac{1}{2}g_D f_{A_1 A_2 A_3} \sum_{i_1,i_2,i_3} g_{i_1 i_2 i_3} \phi^{A_1 i_1} \phi^{A_2 i_2} \phi^{A_3 i_3}, \\
\mathcal{L}_{\phi \phi A}  =& g_D f_{A_1 A_2 A_3} \sum_{i_1, i_2, a_3} g_{a_3 i_1 i_2} \partial^{\mu} \phi^{A_1 i_1} \phi^{A_2 i_2} A_{\mu}^{A_3 a_3}, \\
\mathcal{L}_{\phi \phi A^0}  =& g_d f_{A_1 A_2 A_3} \sum_{i_1} \partial^{\mu} \phi^{A_1}{}_{ i_1}  \phi^{A_2 i_1} A_{\mu}^{A_3 0},
\end{align}
where the cubic couplings $g_{i_1 i_2 i_3}$ and $g_{a_3 i_1 i_2}$ were defined in Eqs.~\eqref{eq:giii} and \eqref{eq:gaii}.

The corresponding vertices 
are given by
\begin{align}
\mathcal{V}(1_\phi^{A_1 i_1}, \, 2_\phi^{A_2 i_2}, \, 3_\phi^{A_3 i_3}) & = - i g_D f^{A_1 A_2 A_3} \left( g_{i_1 i_2 i_3} + g_{i_2 i_3 i_1} + g_{i_3 i_1 i_2} \right), \\
\mathcal{V}(1_{\phi}^{A_1 i_1}, \, 2_{\phi}^{A_2 i_2}, \, 3_{A}^{A_3 a_3})&= g_D f^{A_1 A_2 A_3} g_{a_3 i_1 i_2 }  \epsilon_3 \ccdot (p_1-p_2),\\
\mathcal{V}(1_\phi^{A_1 i_1}, \, 2_\phi^{A_2 i_2}, \, 3_{A}^{A_3 0}) & =  g_d f^{A_1 A_2 A_3} \delta_{i_1 i_2}\epsilon_3 \ccdot \left(p_1 -p_2 \right).
\end{align}

We also need three additional four-point interactions,
\begin{align}
\mathcal{L}_{AAAA_0}  =&- \frac{g_D g_d }{2}  f_{A_1 A_2}{}{}^A f_{ A_3 A_4 A}  \sum_{a_1, a_2, a_3}  g_{a_1 a_2 a_3}  A^{A_1 a_1} \ccdot A^{A_3 a_2}  \left[A^{A_2 a_3} \ccdot A^{A_4 0} +  A^{A_2 0} \ccdot A^{A_4 a_3} \right], \label{eq:LAAAA0} \\
\mathcal{L}_{AAA^0A^0}  =&- \frac{ g_d^2}{2 }  f_{A_1 A_2}{}{}^{A } f_{ A_3 A_4 A} \sum_{a} \left[ A^{A_1}_{a}\ccdot A^{A_3 a} A^{A_2 0} \ccdot A^{A_4 0}+A^{A_1 a } \ccdot A^{A_3 0}\left( A^{A_2}_{a} \ccdot A^{A_4}_{0}+A^{A_2}_{0} \ccdot A^{A_4}_{ a}\right) \right], \label{eq:LAAA0A0} \\
\mathcal{L}_{AA \phi \phi}  =&- \frac{ g_D^2}{2}  f_{A_1 A_2}{}{}^{A } f_{ A_3 A_4 A}   \sum_{a_1, a_2, i_3, i_4}g_{a_1 a_2 i_3 i_4}  A^{A_1 a_1} \ccdot A^{A_3 a_2}\phi^{A_2 i_3} \phi^{A_4 i_4},
\end{align}
where we have defined the additional quartic coupling
\be
g_{a_1 a_2 i_3 i_4} \equiv \int_{\mathcal{N}} \psi_{a_1} \psi_{a_2} Y_{n, i_3}Y^n_{i_4}.
\ee
We do not write explicitly the $\phi^4$, $A_0^3 A$, or $A_0^4$ contact interactions since the corresponding amplitudes automatically have the requisite high-energy behavior.

The corresponding quartic vertices 
are given by
\begin{align}
\mathcal{V}(1_A^{A_1 a_1}, \, 2_A^{A_2 a_2}, \, 3_{A}^{A_3 a_3},\, 4_{A}^{A_40})  =& - i g_D g_d g_{a_1 a_2 a_3} \Big[ \epsilon_1 \ccdot \epsilon_2 \, \epsilon_3 \ccdot \epsilon_4 \left(f^{ A_1 A_3 A} f^{A_2 A_4 A}+f^{A_1 A_4 A} f^{A_2 A_3 A} \right)  \nn \\
&  +\epsilon_1 \ccdot \epsilon_3 \, \epsilon_2 \ccdot \epsilon_4 \left(f^{A_1 A_2 A} f^{A_3 A_4 A}-f^{A_1 A_4 A} f^{ A_2 A_3 A} \right)  \nn \\
& -\epsilon_1 \ccdot \epsilon_4 \, \epsilon_2 \ccdot \epsilon_3 \left(f^{ A_1 A_2 A} f^{A_3 A_4 A}+f^{A_1 A_3 A} f^{ A_2 A_4 A} \right)  \Big], \\
\mathcal{V}(1_A^{A_1 a_1}, \, 2_A^{A_2 a_2}, \, 3_{A}^{A_3 0}, \, 4_{A}^{A_4 0})  =&- i g_d^2 \delta_{a_1 a_2} \Big[ \epsilon_1 \ccdot \epsilon_2 \, \epsilon_3 \ccdot \epsilon_4 \left(f^{ A_1 A_3}{}{}_{A} f^{A_2 A_4 A}+f^{A_1 A_4}{}{}_{A} f^{A_2 A_3 A} \right)  \nn \\
&  +\epsilon_1 \ccdot \epsilon_3 \, \epsilon_2 \ccdot \epsilon_4 \left(f^{A_1 A_2}{}{}_{A} f^{A_3 A_4 A}-f^{A_1 A_4}{}{}_{A} f^{ A_2 A_3 A} \right)  \nn \\
& -\epsilon_1 \ccdot \epsilon_4 \, \epsilon_2 \ccdot \epsilon_3 \left(f^{ A_1 A_2}{}{}_{A} f^{A_3 A_4 A}+f^{A_1 A_3}{}{}_{A} f^{ A_2 A_4 A} \right)  \Big], \\
\mathcal{V}(1_A^{A_1 a_1}, \, 2_A^{A_2 a_2}, \, 3_\phi^{A_3 i_3}, \, 4_\phi^{A_4 i_4})  =&- i g_D^2 g_{a_1 a_2 i_3 i_4} \left(f^{A_1 A_3}{}{}_{A} f^{A_2 A_4 A} + f^{A_1 A_4}{}{}_{A} f^{A_2 A_3 A} \right) \epsilon_1 \ccdot \epsilon_2.
\end{align}

We can now calculate the remaining amplitudes and look for additional sum rules. We find that only the processes 
\be \label{eq:AAAphi}
A_L^{A_1 a_1} A_L^{A_2 a_2} \rightarrow A_L^{A_3 a_3}  \phi^{A_4 i_4},
\ee 
and
\be \label{eq:AAphiphi}
A_L^{A_1 a_1} A_L^{A_2 a_2} \rightarrow \phi^{A_3 i_3}  \phi^{A_4 i_4},
\ee 
give additional sum rules, which arise from setting to zero the $E^2$ and $E^2 \cos \theta$ parts of the amplitudes.

From the process \eqref{eq:AAAphi} we get the sum rule
\be
\sum_a \left( g_{a_1 a_2}{}{}^{a}g_{a a_3 i_4}+  g_{a_2 a_3}{}{}^{a} g_{a a_1 i_4}  +  g_{a_1 a_3}{}{}^{a} g_{a a_2 i_4}\right) =0, \label{eq:cyclicidentity}
\ee
which can be derived from the associativity relation 
\be
\int_{\mathcal{N}} 
\contraction{}{\partial_n  \psi_{(a_1} }{}{ \psi_{a_2)} }
\contraction{\partial_n \psi_{(a_1}  \psi_{a_2)} }{\psi_{a_3} }{}{Y_{n, i_4}}
\partial_n \psi_{(a_1}  \psi_{a_2)} \psi_{a_3} Y_{n, i_4}
=
\int_{\mathcal{N}} 
\contraction[2ex]{}{\partial_n \psi_{(a_1}}{ \psi_{a_2)} \psi_{a_3} }{Y_{n, i_4}}
\contraction{\partial_n \psi_{(a_1} }{ \psi_{a_2) } }{}{ \psi_{a_3}}
\partial_n \psi_{(a_1}  \psi_{a_2)} \psi_{a_3} Y_{n, i_4} \, .
\ee 
The other sum rules from this process can be written as cyclic permutations of
\be \label{eq:AAAphirule}
  \sum_a \left( \lambda_a^{-1}\left(\lambda_{a_1}-\lambda_{ a_2}- \lambda_a \right) g_{a_1 a_2}{}{}^{a} g_{a a_3 i_4} -2g_{a_1 a_3}{}{}^{a} g_{a a_2 i_4}\right)+2 \sum_i g_{a_1 a_2}{}{}^{i} g_{a_3 i_4 i}=0,
\ee
which can be derived from Eq.~\eqref{eq:cyclicidentity} combined with the associativity relation
\begin{align}
\int_{\mathcal{N}}
\contraction{}{\partial_n \psi_{a_1}}{}{ \psi_{a_2}}{}
\contraction{\partial_n \psi_{a_1} \psi_{a_2} }{ \psi_{a_3}}{}{Y_{n, i_4}}
\partial_n \psi_{a_1} \psi_{a_2} \psi_{a_3} Y_{n, i_4}
& =
\int_{\mathcal{N}}
\contraction[2ex]{}{\partial_n \psi_{a_1} }{\psi_{a_2} \psi_{a_3}}{ Y_{n, i_4}}
\contraction{\partial_n \psi_{a_1}}{ \psi_{a_2}}{}{ \psi_{a_3}}
\partial_n \psi_{a_1} \psi_{a_2} \psi_{a_3} Y_{n, i_4} \, .
\label{eq:assoc3}
\end{align}

Similarly, from the process \eqref{eq:AAphiphi} we get the two sum rules
\begin{align}
\sum_i (g_{a_1 i_4}{}{}^i g_{a_2 i_3 i}-g_{a_1 i_3}{}{}^i g_{a_2 i_4 i})+\sum_a \lambda_a^{-1} ( g^a{}_{a_1 i_4} g_{a a_2 i_3}-g^a{}_{a_1 i_3} g_{a a_2 i_4}) &= 0, \\
\sum_i g_{a_1 i_4}{}{}^i g_{a_2 i_3 i}+\sum_a ( \lambda_a^{-1}  g^a{}_{a_1 i_4} g_{a a_2 i_3}-g_{a_1 a_2}{}{}^a g_{a i_3 i_4})-V^{-1} \delta_{a_1 a_2} \delta_{i_3 i_4} &= 0.
\end{align}
These can be derived from the associativity relations
\be
\int_{\mathcal{N}}
\contraction{}{\psi_{a_1}}{}{\psi_{a_2}}
\contraction{\psi_{a_1}\psi_{a_2}}{Y_{n, i_3}}{}{Y_{n, i_4}}
\psi_{a_1} \psi_{a_2} Y_{n, i_3}Y^n_{i_4}
=
\int_{\mathcal{N}}
\contraction{}{\psi_{a_1}}{\psi_{a_2}}{Y_{n, i_3}}
\contraction[2ex]{\psi_{a_1}}{\psi_{a_2}}{Y_{n, i_3}}{Y_{n, i_4}}
\psi_{a_1} \psi_{a_2} Y_{n, i_3}Y^n _{i_4}
=
\int_{\mathcal{N}}
\contraction[2ex]{}{\psi_{a_1}}{\psi_{a_2}Y_{n, i_3}}{Y_{n, i_4}}
\contraction{\psi_{a_1}}{\psi_{a_2}}{}{Y_{n, i_3}}
\psi_{a_1} \psi_{a_2} Y_{n, i_3}Y^n_{i_4} .
\ee

\subsection{Examples}
We now discuss some explicit examples of internal manifolds to see how the sum rules are satisfied in concrete cases.

First we make a comment on the behavior of the sum rules under Weyl transformations, as this will be useful for normalizing the geometric data in what follows.
Suppose we rescale the metric $\gamma_{n_1 n_2}$ of the internal space $\mathcal{N}$ by an overall factor
\be \label{eq:rescale}
\gamma_{n_1 n_2} \rightarrow \Omega^{2/N} \gamma_{n_1 n_2}, 
\ee
where $\Omega>0$ is a constant. Then under this rescaling the various geometric quantities transform as
\begin{align}
V & \rightarrow \Omega V, & \psi_a & \rightarrow \Omega^{-1/2} \psi_a, &  g_{a_1 a_2 a_3} & \rightarrow \Omega^{-1/2} g_{a_1 a_2 a_3}, & g_{a_1 i_2 i_3} & \rightarrow \Omega^{-1/2} g_{a_1 i_2 i_3}, \nn \\
\lambda_a  & \rightarrow \Omega^{-2/N} \lambda_a, & Y_{n, i}  & \rightarrow \Omega^{(2-N)/2N} Y_{n, i},  & g_{a_1 a_2 i_3}&  \rightarrow \Omega^{-(2+N)/2N} g_{a_1 a_2 i_3}, & g_{i_1 i_2 i_3} &  \rightarrow \Omega^{-(2+N)/2N} g_{i_1 i_2 i_3}.
\end{align}
It can be checked that under these transformations the sum rules transform homogeneously in $\Omega$. As a consequence we can normalize $V$ to any given value without loss of generality when checking the sum rules in explicit examples.

\subsubsection{A circle}
\label{sec:circle}

Consider first the example of a circle, $\mathbb{S}^1$, which is the only closed manifold with $N=1$. We set the radius equal to one, so the volume is $V=2 \pi$. The real scalar eigenfunctions $\psi_a$ can be written as even or odd trigonometric functions,
\be
\psi_{2k-1} = \frac{1}{\sqrt{\pi}} \sin(k \theta), \quad \psi_{2k} = \frac{1}{\sqrt{\pi}} \cos(k \theta), \quad \lambda_{2k-1} = \lambda_{2k} = k^2,\quad k\in \mathbb{Z}_{>0}.
\ee
These lead to massive vector multiplets in the dimensionally reduced theory.
The only vector eigenfunction on a circle is the zero mode,
\be
Y_{\theta} = \frac{1}{\sqrt{2\pi}},
\ee
which gives a multiplet of massless scalars in the dimensionally reduced theory. 

We can now find the cubic couplings by evaluating triple overlap integrals of these eigenfunctions. Since the only vector is a zero mode, we always have $ g_{i_1 i_2 a_3}=0$ due to the orthonormality of the scalar eigenfunctions and the constant function. From the definition \eqref{eq:tripleint2}, we have that the only nonzero couplings $g_{a_1 a_2 i}$ take the form
\be
g_{2k_1-1, 2k_1, i} = -g_{2k_1, 2k_1-1, i}= \frac{k_1}{\sqrt{2 \pi}} . 
\ee
For the scalar triple overlap integrals $g_{a_1 a_2 a_3}$, the non-zero cases are
\begin{align}
g_{2k_1, 2k_2, 2k_3}  &= \frac{1}{2\sqrt{\pi}} \left( \delta_{k_1+k_2, k_3} +\delta_{|k_1-k_2|, k_3} \right), \label{eq:eventriple}\\
g_{2k_1, 2k_2-1, 2k_3-1} & = \frac{1}{2\sqrt{\pi}} \left(- \delta_{k_2+k_3, k_1} +\delta_{|k_2-k_3|, k_1} \right), 
\end{align}
plus those related to these by symmetry.

As an example amplitude, consider the scattering of identical external vectors with even KK indices, $a_i=2k_1 $. The nonvanishing cubic couplings imply that in this process we can only exchange gluons or an even massive vector multiplet with twice the mass of the external particle. The corresponding solution to the sum rule in Eq.~\eqref{eq:sumrule1b} is given by
\be
\lambda_{a_1} = k_1^2, \quad \lambda_a = 4k_1^2, \quad  g_{a_1 a_1 a} = \frac{1}{2 \sqrt{\pi}}, \quad V =2 \pi. 
\ee
Similarly, if we consider the inelastic process \eqref{eq:A1A2A1A2} with $a_1=2k_1-1$ and $a_2 = 2k_1$, then we can exchange the multiplet of massless scalars plus an odd vector multiplet with twice the mass of the external particles. The corresponding solution to the sum rule in Eq.~\eqref{eq:sumrule1} is given by
\be
\lambda_{a_1} = \lambda_{a_2} = k_1^2, \quad \lambda_a = 4k_1^2, \quad g_{a_1 a_2 a} = \frac{1}{2 \sqrt{\pi}}, \quad g_{a_1 a_2 i} = \frac{k_1}{\sqrt{2 \pi}} .
\ee

\subsubsection{Tori}
\label{sec:tori1}
As our next example, consider the square flat $N$-torus, $\mathbb{T}^N=\mathbb{R}^N/(2 \pi \mathbb{Z})^N$, which is the product of $N$ circles with unit radii. The metric is
\be
ds^2_{\mathbb{T}^N} = \sum_{j=1}^N d\theta_j^2,
\ee
where $0 \leq \theta_j < 2 \pi$.
The scalar eigenfunctions are products of the eigenfunctions of the circle, so we can label them by an $N$-vector $\vec{a}_1=(a_{11}, \dots, a_{1N})$,
\be
\psi_{\vec{a}_1}  =\psi_{a_{11}}\psi_{a_{12}} \dots \psi_{a_{1N}}, \quad \lambda_{\vec{a}_1} = \sum_{j=1}^N \lambda_{a_{1j}} ,
\ee
where $\psi_{a_{1j}}$ is a function of $\theta_j$. 
The scalar triple overlap integrals are just products of those of the circle, 
\be \label{eq:tripleprod}
g_{\vec{a}_1 \vec{a}_2 \vec{a}_3} = \prod_{j=1}^N g_{a_{1 j} a_{2 j} a_{3 j}}.
\ee
There are also vector eigenfunctions, but we will only consider an amplitude for which they can be ignored.

Consider then the amplitude with identical external particles described by a product of even functions, $\vec{a}_1 = (2 k_1, 2 k_2, \dots, 2k_n) $ with $k_i \in \mathbb{Z}_{>0}$. From Eqs.~\eqref{eq:eventriple} and \eqref{eq:tripleprod}, the nonvanishing triple overlap integrals are labeled by nonempty elements $r$ of the powerset (i.e. the set of subsets) of $S={\{1, \, \dots, \, N\}}$,
\be \label{eq:torusdata}
g_{\vec{a}_1 \vec{a}_1 \vec{a}_r} = \left( \frac{1}{2 \sqrt{\pi}}\right)^{|r|}\left( \frac{1}{\sqrt{2 \pi}}\right)^{N-|r|}, \quad a_{rj} =
 \begin{cases}
4k_j \quad {\rm if} \quad j \in r \\
0 \quad {\rm otherwise}
\end{cases}
, \quad r \in 2^S \setminus \{\} .
\ee
This implies that the amplitude involves the exchange of gluons and $2^N-1$ different massive vector multiplets. The sum rule in Eq.~\eqref{eq:sumrule1b} then becomes 
\be
\sum_{r \in 2^S\setminus \{\} } \left( 4 \lambda_{\vec{a}_1} - 3 \lambda_{\vec{a}_r} \right) g^2_{\vec{a}_1 \vec{a}_1 \vec{a}_r} + 4 V^{-1}  \lambda_{\vec{a}_1} =0 \, ,
\ee
which we can explicitly verify by substituting the values in Eq.~\eqref{eq:torusdata}.

\subsubsection{A sphere}
As a final example for this section, consider the two-sphere, $\mathbb{S}^2$, with unit radius and the round metric,
\be
ds^2_{\mathbb{S}^2} = d \theta^2 + \sin^2 \theta d \phi^2,
\ee 
where $0 \leq\theta \leq \pi$ and $0 \leq \phi < 2 \pi$.

The non-constant eigenfunctions $\psi_a$ of the scalar Laplacian are labeled by two indices: $a = (l, m)$, where $l \in \mathbb{Z}_{> 0}$ and $m \in \{-l, -l+1, \ldots, l-1, l\}$. A real orthonormal basis is given by the real spherical harmonics,
\be
\psi_{lm} \equiv 
\begin{cases}
\frac{1}{\sqrt{2}} \left( Y_l{}^{-m}+ (-1)^m Y_l{}^m \right) \quad & {\rm if} \quad m>0,  \\
Y_{l}{}^0 \quad & {\rm if} \quad m=0,\\
\frac{i}{\sqrt{2}} \left( Y_l{}^{m}- (-1)^m Y_l{}^{-m} \right) \quad &  {\rm if} \quad m<0,
\end{cases}
\ee
where $Y_l{}^m$ are the usual complex spherical harmonics. The corresponding eigenvalues are
\be
\lambda_{l m} = l(l+1).
\ee

The orthonormal and transverse eigenfunctions of the vector Laplacian are labeled by $i=(\tilde{l}, \tilde{m})$ with $\tilde{l} \in \mathbb{Z}_{> 0}$ and $\tilde{m} \in \{-\tilde{l}, -\tilde{l}+1, \ldots, \tilde{l}-1, \tilde{l} \}$ and can be written as~\cite{Higuchi:1986wu}
\be
Y_{n_1, \, \tilde{l}\tilde{m}} = \frac{1}{\sqrt{\tilde{l}(\tilde{l}+1)}}\epsilon_{n_1 n_2} \partial^{n_2} \psi_{\tilde{l}\tilde{m}},
\ee
where $\epsilon_{n_1 n_2}$ is the antisymmetric tensor with components
\be
\epsilon_{\theta \theta}=\epsilon_{\phi \phi} =0, \quad \epsilon_{\theta \phi}=-\epsilon_{\phi \theta} = \sin \theta.
\ee
In components, the transverse vector eigenfunctions are given by
\be
\begin{pmatrix}
Y_{ \theta, \, \tilde{l}\tilde{m}} \\
Y_{ \phi, \, \tilde{l}\tilde{m}}
\end{pmatrix} =
\frac{1}{\sin \theta \sqrt{\tilde{l}(\tilde{l}+1)}} \begin{pmatrix}
\partial_{\phi} \psi_{\tilde{l}\tilde{m}}\\
-\sin^2 \theta \, \partial_{\theta} \psi_{\tilde{l}\tilde{m}}
\end{pmatrix},
\ee
and their eigenvalues are
\be
\lambda_{\tilde{l}\tilde{m}} = \tilde{l}(\tilde{l}+1).
\ee

The scalar triple overlap integrals are defined as
\begin{align}
g_{a_1 a_2 a_3} =  \int_{\mathbb{S}^2} \psi_{l_1 m_1} \psi_{l_2 m_2} \psi_{l_3 m_3},
\end{align}
where $a_j = (l_j, m_j)$.
A useful identity for evaluating these is
\be \label{eq:CGidentity}
 \int_{\mathbb{S}^2}   Y_{l_1}{}^{m_1}  Y_{l_2}{}^{m_2}  Y_{l_3}{}^{-m_3}  = (-1)^{m_3} \sqrt{ \frac{(2 l_1 + 1) (2 l_2 + 1)}{4 \pi (2 l_3 + 1)}} C^{l_1 l_2 l_3}_{000} C^{l_1 l_2 l_3}_{m_1 m_2 m_3},
\ee
where $C^{l_1 l_2 l_3}_{m_1 m_2 m_3}$ are Clebsch--Gordan coefficients. 
Similarly, the triple overlap integrals with one vector eigenfunction labeled by $i_3=(\tilde{l}_3, \tilde{m}_3)$ are given by
\be
g_{a_1 a_2 i_3}= \frac{1}{2 \sqrt{\tilde{l}_3(\tilde{l}_3+1)}} \int_{\mathbb{S}^2} \left[  \frac{\psi_{l_2 m_2}}{\sin \theta}\left( \partial_{\theta} \psi_{l_1 m_1} \partial_{\phi} \psi_{\tilde{l}_3 \tilde{m}_3} -\partial_{\phi} \psi_{l_1 m_1} \partial_{\theta} \psi_{\tilde{l}_3 \tilde{m}_3} \right) - \left( l_1 \leftrightarrow l_2 \right) \right] .
\ee

Consider now the scattering of identical massive vectors with $a_1=(l_1, l_1)$, so that $\lambda_{a_1} = l_1 (l_1 +1)$. If we write $a=(2 l ,m)$, then $\lambda_a = 2l(2l+1)$ and the nonvanishing scalar triple overlap integrals are
\begin{align}
g_{a_1 a_1 a} & = 
\frac{ \delta_{m, 0} (-1)^{l_1} \sqrt{4 l+1} \Gamma \! \left(l+\frac{1}{2}\right) \Gamma \! \left(2 l_1+2\right)}{4^{l+1}l! \left(2 l_1-2 l\right)!\,  \Gamma \! \left(l-l_1+\frac{1}{2}\right)
\Gamma \! \left(l+l_1+\frac{3}{2}\right)}+  \frac{\delta_{m, 2 l_1} \Gamma \left(l+\frac{3}{2}\right) \sqrt{ \Gamma (2 l+1)}}{2 \pi ^{3/4} \Gamma (l+1) \sqrt{ \Gamma \left(2
   l+\frac{3}{2}\right)}} \label{eq:S2triple},
\end{align}
where $l \in \mathbb{Z}_{>0}$ and $1\leq l \leq l_1$.
So in this amplitude we can exchange gluons, multiplets of massive vectors with $a = (2l,0)$ for $l=1, \ldots, l_1$, and a multiplet of massive vectors with $a=(2l_1, 2l_1)$.
Plugging everything into the sum rule in Eq.~\eqref{eq:sumrule1b} with $V= 4\pi$, we can verify that this gives a very nontrivial solution. 

For a simple example with scalar exchange, consider the process \eqref{eq:A1A2A1A2} with $a_1 = (2, -1)$ and $a_2 =(1, 1)$. This involves the exchange of one vector multiplet and two scalar multiplets with the masses and cubic couplings given by
\begin{align}
(a, \lambda_a, g_{a_1 a_2 a}) & \in \left\{ \left( (3, -2), 12, \frac{1}{2} \sqrt{\frac{3}{7 \pi}}\right)\right\},  \\
(i, \lambda_i, g_{a_1 a_2 i} ) & \in \left\{ \left((2, 0), 6,\frac{1}{2}\sqrt{\frac{3}{2 \pi}}\right), \left((2, 2), 6, \frac{1}{2 \sqrt{2 \pi}}\right) \right\} .
\end{align}
Plugging these  into the sum rule in Eq.~\eqref{eq:sumrule1}, we can verify that they indeed provide a solution.

\section{General relativity}
\label{sec:GR}

In this section we consider amplitudes in a general dimensional reduction of pure GR on a closed internal manifold down to a lower-dimensional flat spacetime. The unitarity of four-dimensional amplitudes coming from a five-dimensional compactification has been studied in Refs.~\cite{Schwartz:2003vj, Chivukula:2019rij, Chivukula:2019zkt}.

\subsection{Higher-dimensional theory}

We start with the $D$-dimensional Einstein--Hilbert Lagrangian,
\be
\mathcal{L} = \frac{ M_{D}^{D-2}}{2} \sqrt{-G} R(G),
\ee
where $G_{A_1 A_2}$ is the  $D$-dimensional metric, with determinant $G$, and $M_{D}$ is the $D$-dimensional reduced Planck mass.

\subsubsection{Background and fluctuations}
Now consider a background metric $\bar{G}_{A_1 A_2}$ that is a product space $\mathcal{M} \times \mathcal{N}$ of flat $d$-dimensional Minkowski space $\mathcal{M}$ and a closed, smooth, connected, orientable $N$-dimensional Riemannian manifold $\mathcal{N}$.
The full metric is a product given by
\be
ds^2 = \bar{G}_{A B} dX^A dX^B= \eta_{\mu \nu} dx^{\mu} dx^{\nu} + \gamma_{mn} dy^m dy^n \, ,
\ee
where $x^{\mu}$ are coordinates on $\mathcal{M}$ and $y^m$ are coordinates on $\mathcal{N}$.

For this metric to solve the higher-dimensional vacuum Einstein equations, $\mathcal{N}$ must be a Ricci-flat manifold,
\be R_{mn}=0.\label{ricciflatconde}\ee
 Compact Ricci-flat manifolds seem to be relatively rare Riemannian manifolds \cite{Besse:1987pua}. The known examples are flat tori, Calabi--Yau manifolds \cite{Yau1978}, compact $G_2$ and ${\rm Spin}(7)$ manifolds \cite{joyce2000a}, and their products.\footnote{These manifolds all have special holonomy and it is an open problem whether or not there exists a simply connected closed Ricci-flat manifold with full $SO(N)$ holonomy.}

We expand the full metric perturbatively around the background metric $\bar{G}_{A_1 A_2}$,
\be
G_{A_1 A_2} = \bar{G}_{A_1 A_2} + \frac{2}{M_{D}^{\frac{D-2}{2}}} \delta G_{A_1 A_2},
\ee
where the perturbation is given by
\be
 \delta G_{A_1 A_2} = H_{A_1 A_2} + \frac{1}{M_{D}^{\frac{D-2}{2}}} \left( H^2_{A_1 A_2} - \frac{1}{2} H H_{A_1 A_2} + \frac{1}{4(D-2)}\bar{G}_{A_1 A_2} H^2  \right) + \frac{1}{M_D^{D-2}} H^3_{A_1 A_2} \, .
\ee
The slightly unusual field basis $H_{A_1 A_2}$ is chosen to minimize the number of interaction terms at cubic and quartic order (without any gauge fixing and with the usual spin-2 kinetic term). 

Expanding the Lagrangian up to quartic order gives 
\begin{align}
{1\over \sqrt{-\rule{0pt}{2ex}\bar{G}}}\mathcal{L}_{(2)} = &- \frac{1}{2} (\nabla_{A_1} H_{A_2 A_3} )^2 + \nabla_{A_1} H_{A_2 A_3} \nabla^{A_2} H^{A_1 A_3}- \nabla_{A_1} H \nabla_{A_2} H^{A_1 A_2}+\frac{1}{2} (\nabla_{A_1} H)^2, \label{eq:GR2} \\
{1\over \sqrt{-\rule{0pt}{2ex}\bar{G}}}\mathcal{L}_{(3)} = & \frac{1}{M_D^{\frac{D-2}{2}}}  H^{A_1 A_2}  \Big[ \nabla_{A_1} H_{A_3 A_4} \nabla_{A_2} H^{A_3 A_4} -2  \nabla_{A_1} H^{A_3 A_4} \nabla_{A_3} H_{A_2 A_4}+  \nabla_{A_1} H \nabla^{A_3} H_{A_2 A_3}  \nn \\
& +  \frac{1}{2} \nabla_{A_3} H \nabla^{A_3} H_{A_1 A_2} - \frac{1}{2} \nabla_{A_1} H \nabla_{A_2} H \Big], \label{eq:GR3} \\ 
{1\over \sqrt{-\rule{0pt}{2ex}\bar{G}}}\mathcal{L}_{(4)} =&\frac{1}{M_D^{D-2}}\Big[  H^{A_1 A_2} H^{A_3 A_4}\nabla_{A_1} H_{A_3 A_5} \nabla_{A_4} H_{A_2}{}^{A_5}+(H^2)^{A_1 A_2} \big( \nabla_{A_3} H_{A_1 A_4} \nabla^{A_4} H_{A_2}{}^{A_3}  \nn \\ 
                  \qquad & -  \nabla_{A_1} H_{A_3 A_4} \nabla_{A_2} H^{A_3 A_4}+2 \nabla_{A_1} H^{A_3 A_4} \nabla_{A_3} H_{A_2 A_4} - \nabla_{A_3} H_{A_1 A_4} \nabla^{A_3} H_{A_2}{}^{A_4} \big) + \cdots \Big] \, , \label{eq:GR4}
\end{align}
where in $\mathcal{L}_{(4)}$ we do not show explicitly terms containing the trace or divergence of $H_{A_1 A_2}$, as these will not affect the on-shell four-point amplitudes.
The theory is invariant under diffeomorphisms, which at leading order in the fields act on $H_{A_1 A_2}$ as
\be
\delta H_{A_1 A_2} = \nabla_{A_1} \Xi_{A_2}+\nabla_{A_2} \Xi_{A_1}+\cdots
\ee
where $\Xi_{A_1}$ is a vector gauge parameter.

\subsubsection{Hodge and eigenfunction decompositions}
We now expand $H_{A_1 A_2}$ using the Hodge decomposition, suitably generalized to symmetric tensors \cite{Ishibashi:2004wx, Hinterbichler:2013kwa}, combined with the eigenfunction decomposition. This takes the following form  \cite{Hinterbichler:2013kwa}:
\begin{align}
H_{\mu \nu}(x,y) =&  \sum_a h_{\mu \nu}^a(x) \psi_a(y) + \frac{1}{\sqrt{V}} h_{\mu \nu}^0(x)\, , \\
H_{\mu n}(x,y)  = & \sum_i A_{\mu}^i (x) Y_{n, i}(y)+\sum_a A_{\mu}^a(x) \partial_n \psi_a(y) \, ,\\
H_{mn}(x,y)  = & \sum_{i \notin I_{\rm{Killing}}} \phi^i(x) \left( \nabla_m Y_{n, i}(y) + \nabla_n Y_{m, i}(y) \right) +\sum_{a \notin I_{\rm conf.} } \tilde{\phi}^a(x) \left( \nabla_m \nabla_n \psi_a(y) -  \frac{1}{N} \nabla^2 \psi_a(y) \gamma_{mn} \right)\, \nn \\
&  + \sum_a \frac{1}{N} \phi^a(x) \psi_a(y) \gamma_{mn} +\frac{1}{N} \frac{1}{\sqrt{V}} \phi^0(x) \gamma_{mn}+  \sum_{\mathcal{I}} \phi^{\mathcal{I}}(x) h_{mn, \mathcal{I}}^{TT}(y)\,. \label{eq:hodgeGR}
\end{align}

The new parts in this decomposition compared to the YM case of Section \ref{sec:YMHodge} are as follows.
There is the collection of fields $h_{mn, \mathcal{I}}^{TT} $, which parameterize the transverse traceless modes of the components of the graviton in the internal space. These fields form a basis of transverse, traceless real symmetric tensors on $\mathcal{N}$ and are eigenmodes of the Lichnerowicz operator $\Delta_L$ with eigenvalues $\lambda_{\mathcal{I}}$ \cite{Lichnerowicz},
\be
\Delta_{L} h_{mn, \mathcal{I}}^{TT} \equiv  -\Box h_{mn, \mathcal{I}}^{TT} + \frac{2 R_{(N)}}{N} h_{mn, \mathcal{I}}^{TT}- 2 R_{mpnq} h^{pq \; TT}_{\mathcal{I}}= \lambda_{\mathcal{I}} h_{mn, \mathcal{I}}^{TT},
\ee
\be \nabla^m h_{mn, \mathcal{I}}^{TT}=\gamma^{mn}h_{mn, \mathcal{I}}^{TT}=0,\ee
where in our case the Ricci scalar of the internal space $R_{(N)}$ vanishes due to the Ricci-flat condition.
They also satisfy an orthonormality condition,
\be
\intN h_{mn , \mathcal{I}_1}^{TT}h^{mn, TT}_{\mathcal{I}_2} = \delta_{\mathcal{I}_1 \mathcal{I}_2}.
\ee
The sum over transverse vector eigenfunctions in Eq.~\eqref{eq:hodgeGR} does not include the Killing vectors of $\mathcal{N}$, which are indexed by the set $I_{\rm{Killing}}$, since by definition these satisfy Killing's equation,
\be
\nabla_{(m} Y_{n)}^{i} =0, \quad i \in I_{\rm{Killing}}.
\ee
Similarly, the sum over scalar eigenfunctions in the second term in Eq.~\eqref{eq:hodgeGR} excludes eigenfunctions that are conformal scalars, i.e. scalars that satisfy the equation
\be
\nabla_{m} \nabla_n \psi_a - \frac{1}{N} \Box \psi_a \gamma_{mn} =0,
\ee
which we index by the set $I_{\rm conf.}$.
By Obata's theorem \cite{obata1962}, conformal scalars only exist on manifolds that are isometric to the round $N$-sphere, $\mathbb{S}^N$, and for $N>1$ they correspond to the $l=1$ spherical harmonics, which have eigenvalues saturating the Lichnerowicz bound \eqref{eq:lich2}. For $N=1$, all non-constant scalars are conformal scalars.

Note that the Lichnerowicz eigenvalues $\lambda_{\mathcal{I}}$ can be negative on general Einstein manifolds. Such a negative eigenvalue would lead to a tachyonic scalar field in the dimensionally-reduced theory in the Ricci-flat case of interest to us, as we show below. However, the existence of a nonzero parallel spinor on a cover of a closed Ricci-flat manifold guarantees that $\lambda_{\mathcal{I}} \geq 0$, since on such manifolds the Lichnerowicz operator can be related to the square of the Rarita--Schwinger operator \cite{Wang91, Gibbons:2002th, Dai2005}. Since all known closed Ricci-flat manifolds have covers with special holonomy, which thus admit parallel spinors, they are all stable in this sense.\footnote{For positively curved Einstein manifolds there are known compact examples with negative Lichnerowicz eigenvalues, such as the infinite number of B\"ohm metrics defined on low-dimensional spheres and products of low-dimensional spheres \cite{Bohm1998, Gibbons:2002th}, but these cannot be used to compactify pure gravity to flat space since they are not Ricci-flat.}

\subsubsection{Gauge fixing}
We can similarly decompose the vector gauge parameter $\Xi_{A_1}$,
\begin{align}
\Xi_{\mu} & = \sum_a \xi_{\mu}^a \psi_a + \frac{1}{\sqrt{V}} \xi_{\mu}^0\, ,\nn \\
\Xi _n & =  \sum_i \xi^i Y_{n, i}+\sum_a \xi^a \partial_n \psi_a \, .
\end{align}
At leading order in the fields, the $d$-dimensional fields $A_{\mu}^a$, $ \phi^i$, and $\tilde{\phi}^a$ transform under this gauge symmetry as
\begin{align} \label{eq:GRstuck}
\delta A_{\mu}^a = \xi_{\mu}^a + \partial_{\mu} \xi^a+\cdots, \quad \delta \phi^i = \xi^i+\cdots, \quad \delta \tilde{\phi}^a = 2\xi^a+\cdots. 
\end{align}
We can thus use $\xi_{\mu}^a$, $\xi^i$ and $\xi^a$ to eliminate these fields by choosing the gauge
\be \label{eq:gf}
 A_{\mu}^a=0, \quad \phi^i=0, \quad \tilde{\phi}^a = 0.
\ee
This algebraic gauge fixing can be used to determine the gauge parameters appearing in \eqref{eq:GRstuck} order-by-order in an expansion in powers of the fields. When $N=1$ there are no $\tilde{\phi}^a$ fields, since every scalar is a conformal scalar, and we can instead use the gauge freedom to eliminate the scalars $\phi^a$. 

The remaining gauge transformations are parameterized by the zero mode vector $\xi^0_{\mu}$ and the scalars $\xi_{i}$ with  $i \in I_{\rm{Killing}}$. At leading order, the first of these acts on the zero mode $h^0_{\mu \nu}$ as
\be
\delta h^0_{\mu \nu} = \partial_{\mu} \xi^0_{\nu}+\partial_{\nu} \xi^0_{\mu}+\cdots,
\ee
which is the linearized $d$-dimensional diffeomorphisms acting on the graviton. The gauge parameters $\xi_{i}$  with  $i \in I_{\rm{Killing}}$ are the gauge parameters for the massless vectors corresponding to the Killing vectors of $\mathcal{N}$.

\subsection{Lower-dimensional interactions}
Using the decomposition \eqref{eq:hodgeGR} with the gauge fixing \eqref{eq:gf}, we can now integrate over the internal manifold in the higher-dimensional action to obtain the $d$-dimensional Lagrangian. 

\subsubsection{Spectrum}
The lower-dimensional quadratic Lagrangian is obtained by dimensionally reducing \eqref{eq:GR2}.  It contains mixing terms between the tensors and scalars that can be undone with the following transformations:
\begin{align}
h_{\mu \nu}^a & \rightarrow h_{\mu \nu}^a - \frac{(N-1)}{N(d-1)} \eta_{\mu \nu} \phi^a+ \frac{1}{\lambda_a}\frac{D-2}{N(d-1)} \partial_{\mu} \partial_{\nu} \phi^a, \\
h_{\mu \nu}^0 & \rightarrow h_{\mu \nu}^0 -\frac{1}{d-2} \eta_{\mu \nu} \phi^0.
\end{align}
The diagonalized quadratic Lagrangian is then given by
\be
\mathcal{L}_{(2)}  =\mathcal{L}_{(2)}^0 +  \mathcal{L}_{(2)}^a +\mathcal{L}_{(2)}^i + \mathcal{L}_{(2)}^{\mathcal{I}},
\ee
where the different parts are
\begin{align}
\mathcal{L}^0_{(2)} =& - \frac{1}{2} \partial_{\lambda} h^0_{\mu \nu} \partial^{\lambda} h^{0 \mu \nu} + \partial_{\lambda} h^0_{\mu \nu} \partial^{\mu} h^{0\lambda \nu}- \partial_{\mu} h^0 \partial_{\nu} h^{0\mu \nu}+\frac{1}{2} \partial_{\mu} h^0 \partial^{\mu} h^0 \nn \\
& -\frac{(D-2)}{2N(d-2)}\partial_{\mu} \phi^0 \partial^{\mu} \phi^0 , \\
\mathcal{L}^a_{(2)}  =& - \frac{1}{2} \sum_a \left( \partial_{\lambda} h^a_{\mu \nu} \partial^{\lambda} h_a^{\mu \nu} -2 \partial_{\lambda} h^a_{\mu \nu} \partial^{\mu} h_a^{\lambda \nu}+2 \partial_{\mu} h^a \partial_{\nu} h_a^{\mu \nu}- \partial_{\mu} h^a \partial^{\mu} h_a  + \lambda_a \left(h_{\mu \nu}^a h^{\mu \nu}_a -h^a h_a \right) \right)\nn \\
& - \frac{(N-1)(D-2)}{2N^2(d-1)} \sum_a \left( \partial_{\mu} \phi^a \partial^{\mu} \phi_a + \lambda_a \phi^a \phi_a \right),\\
\mathcal{L}^i_{(2)}  =& -\frac{1}{2} \sum_i \left( F_{\mu \nu}^i F^{\mu \nu}_i +2 \lambda_i A_{\mu}^i A^{\mu}_i \right), \\
\mathcal{L}^{\mathcal{I}}_{(2)}  =&  -\frac{1}{2} \sum_{\mathcal{I}} \left( \partial_{\mu} \phi^{\mathcal{I}} \partial^{\mu} \phi_{\mathcal{I}} + \lambda_{\mathcal{I}}\phi^{\mathcal{I}} \phi_{\mathcal{I}} \right).
\end{align}
The spectrum of the dimensionally reduced theory can now be easily read off.  It consists of the following particles:
\begin{enumerate}
\item A massless graviton, $h_{\mu \nu}^0$.
\item A tower of massive spin-2 particles $h_{\mu \nu}^a$ with squared masses $m^2_a = \lambda_a$, one for every non-constant eigenmode of the scalar Laplacian on $\mathcal{N}$.
\item A tower of vectors $A_{\mu}^i$ with squared masses $m^2_i = \lambda_i$, one for each transverse eigenmode of the vector Laplacian on $\mathcal{N}$, including a massless vector for each Killing vector.
\item A massless scalar $\phi^0$, which is the volume modulus.
\item If $N>1$, a tower of massive scalars $\phi^a$ with squared masses $m^2_a = \lambda_a$, one for every non-constant eigenmode of the scalar Laplacian on $\mathcal{N}$.
\item Another tower of massive scalars $\phi^{\mathcal{I}}$ with squared masses $m^2_{\mathcal{I}} = \lambda_{\mathcal{I}}$, one for each eigentensor of the Lichnerowicz operator on $\mathcal{N}$, including a massless scalar for each eigentensor corresponding to a direction in the moduli space of volume-preserving Einstein structures on $\mathcal{N}$.
\end{enumerate}
Note that from the normalization of the graviton we can obtain the usual relation between the $d$-dimensional Planck mass $M_d$ and the $D$-dimensional Planck mass $M_D$,
\be
M_d^{d-2} = V M_D^{D-2}.\label{Planckmasslhre}
\ee

\subsubsection{Cubic interactions}
We can similarly find the cubic interactions for these fields by dimensionally reducing the higher-dimensional cubic interactions, given in Eq.~\eqref{eq:GR3}. For the four-point amplitude with external massive spin-2 KK modes that we are interested in, we only need the cubic interactions involving at least two massive spin-2 particles. These enter the exchange diagrams as shown in Figure \ref{figure1}.

The interactions with three massive spin-2 fields are
\be
\mathcal{L}_{hhh} = \frac{1}{M_D^{\frac{D-2}{2}}} \sum_{a_1, a_2, a_3} g_{a_1 a_2 a_3}  h^{a_1}_{\mu \nu}\Big[ \partial^{\mu} h^{a_2}_{\lambda \rho} \partial^{\nu} h^{a_3 \lambda \rho}-2 \partial^{\mu} h^{a_2}_{\lambda \rho} \partial^{\lambda} h^{a_3 \nu \rho}+ \frac{1}{4} (\lambda_{a_2}-\Box)h^{a_2} h^{a_3 \mu \nu}\Big]+ \cdots \, ,
\ee
where the triple overlap integral $g_{a_1 a_2 a_3}$ is defined in Eq.~\eqref{eq:tripleint1} and
where here and below we only show explicitly the terms that can contribute to the four-point amplitude with massive spin-2 external states. The interactions involving two massive spin-2 particles and the graviton are
\begin{align}
\mathcal{L}_{hhh^0} = &  \frac{1}{M_d^{\frac{d-2}{2}}} h^0_{\mu \nu} \sum_a \left[ \partial^{\mu}h^{a}_{\lambda \rho} \partial^{\nu} h_a^{\lambda \rho} -2 h^a_{\lambda \rho} \partial^{\lambda}\partial^{\rho} h_a^{\mu \nu} +2 h^a_{\lambda \rho} \partial^{\mu} \partial^{\lambda} h_a^{\nu \rho}-2 \partial^{\mu} h^a_{\lambda \rho} \partial^{\lambda} h_a^{\nu \rho}+2 \partial_{\lambda} h^{a \mu \rho} \partial_{\rho} h_a^{\nu \lambda}\right] \nn \\
& - \frac{1}{2M_d^{\frac{d-2}{2}}} h^0 \sum_a \left[ h^a_{\mu \nu} \Box h_a^{\mu \nu}+ \partial_{\lambda} h^a_{\mu \nu} \partial^{\lambda} h_a^{\mu \nu}  \right]+ \cdots ,
\end{align}
where we have used Eq.~\eqref{Planckmasslhre}.
The interactions with two massive spin-2 particles and a vector are
\be
\mathcal{L}_{hhA} =  \frac{2}{M_D^{\frac{D-2}{2}}} \sum_{a_1, a_2, i_3}g_{a_1 a_2 i_3}   h^{a_1}_{\nu \lambda} \left[ \partial^{\mu} h^{a_2 \nu \lambda}-2\partial^{\nu} h^{a_2 \mu \lambda} \right] A^{i_3}_{\mu}+\cdots,
\ee
where the triple overlap integral $g_{a_1 a_2 i_3}$ is defined in Eq.~\eqref{eq:tripleint2}.
Lastly, the interactions between two massive spin-2 particles and the various scalars after diagonalizing the kinetic terms are
\begin{align}
\mathcal{L}_{hh\phi} = & - \frac{(D-2)}{4 N (d-1)M_D^{\frac{D-2}{2}}}\sum_{a_1, a_2, a_3} \left(\lambda_{a_3}-2(\lambda_{a_1}+\lambda_{a_2})+\lambda_{a_3}^{-1}(\lambda_{a_1}-\lambda_{a_2})^2\right)g_{a_1 a_2 a_3}  h^{a_1}_{\mu \nu} h^{a_2 \mu \nu}\phi^{a_3}+\cdots, \\
\mathcal{L}_{hh\phi^0} = & \frac{1}{M_d^{\frac{d-2}{2}}} \frac{(D-2)}{ N (d-2)} \phi^0 \sum_a \lambda_{a}  h^{a}_{\mu \nu} h^{\mu \nu}_a +\cdots, \\
\mathcal{L}_{hh\phi^{\mathcal{I}}} = & \frac{1}{M_D^{\frac{D-2}{2}}} \sum_{a_1, a_2, \mathcal{I}_3} g_{a_1 a_2 \mathcal{I}_3} h_{\mu \nu}^{a_1} h^{a_2 \mu \nu} \phi^{\mathcal{I}_3}+\cdots , 
\end{align}
where we have defined the new triple overlap integral
\be
g_{a_1 a_2 \mathcal{I}_3} \equiv \int_{\mathcal{N}}  \partial_n \psi_{a_1} \partial_m \psi_{a_2} h^{mn}_{TT, \mathcal{I}_3},
\ee
which is symmetric in its first two indices.

We now vary the interactions to find the corresponding vertices. The vertex for three massive spin-2 particles is given by
\begin{align}
\mathcal{V}(1_h^{a_1}, \, 2_h^{a_2},\, 3_h^{a_3}) = &-\frac{i g_{a_1 a_2 a_3}}{4M_D^{\frac{D-2}{2}}} \epsilon_2 \ccdot \epsilon_3 \left[4 \epsilon_1 \ccdot p_3(\epsilon_2 \ccdot \epsilon_3 \, \epsilon_1 \ccdot p_2 -2 \epsilon_1 \ccdot \epsilon_2 \, \epsilon_3 \ccdot p_2) +(2 p_1\ccdot p_2-\lambda_{a_1}) \, \epsilon_1 \ccdot \epsilon_1 \, \epsilon_2 \ccdot \epsilon_3\right] \nn \\
&+ {\rm \, five \, \, \, permutations},
\end{align}
where here and below we write the polarization tensors as products of null vectors, $\epsilon_{\mu \nu} = \epsilon_{\mu} \epsilon_{\nu}$, and we put the first two legs on shell, since this is all we need to compute the four-point massive spin-2 amplitudes.
The fully on-shell vertex is
\be
\mathcal{V}(1_h^{a_1},\, 2_h^{a_2}, \, 3_h^{a_3})\Big|_{\rm on \,\,shell} =\frac{2i }{M_D^{\frac{D-2}{2}}}  g_{a_1 a_2 a_3} \left( \epsilon_1 \ccdot \epsilon_2 \, \epsilon_3 \ccdot p_1 + \epsilon_2 \ccdot \epsilon_3 \, \epsilon_1 \ccdot p_2+\epsilon_1 \ccdot \epsilon_3 \, \epsilon_2 \ccdot p_3 \right)^2,
\ee
which is the same cubic tensor structure as for the graviton in GR.  {

The vertex describing the gravitational coupling of each massive spin-2 particle is
\begin{align}
\mathcal{V}(1_h^{a_1}, \, 2_h^{a_2}, \, 3_h^0)  =&  \frac{i}{4M_d^{\frac{d-2}{2}}} \delta_{a_1 a_2} \Big[8(\epsilon_2 \ccdot \epsilon_3 \, \epsilon_1 \ccdot p_2 -\epsilon_1 \ccdot \epsilon_3 \, \epsilon_2 \ccdot p_1)(\epsilon_2 \ccdot \epsilon_3 \, \epsilon_1 \ccdot p_2 +\epsilon_1 \ccdot \epsilon_2 \, \epsilon_3 \ccdot p_1) \nn\\
& +(\epsilon_1 \ccdot \epsilon_2)^2(p_3 \ccdot p_3 \, \epsilon_3 \ccdot \epsilon_3 -4 \epsilon_3 \ccdot p_1 \, \epsilon_3 \ccdot p_2) \Big] + (1 \leftrightarrow 2) \, . 
\end{align}
This reduces on-shell to 
\be
\mathcal{V}(1_h^{a_1},\, 2_h^{a_2}, \, 3_h^0)\Big|_{\rm on \,\, shell} =\frac{2i }{M_d^{\frac{d-2}{2}}} \delta_{a_1 a_2} \left( \epsilon_1 \ccdot \epsilon_2 \, \epsilon_3 \ccdot p_1 + \epsilon_2 \ccdot \epsilon_3 \, \epsilon_1 \ccdot p_2+\epsilon_1 \ccdot \epsilon_3 \, \epsilon_2 \ccdot p_3 \right)^2,
\ee
which is gauge invariant and includes the universal minimal coupling interactions mandated by the $S$-matrix equivalence principle \cite{Weinberg:1964ew}.  Note that the cubic interactions among the massive and massless spin-2 fields are all of the type given by requiring asymptotic subluminality in eikonal scattering \cite{Camanho:2014apa, Hinterbichler:2017qyt,Bonifacio:2017nnt}.}

The vertex for two massive spin-2 particles and a spin-1 particle is
\be
\mathcal{V}(1_h^{a_1}, \, 2_h^{a_2}, \, 3_A^{i_3})  = \frac{\sqrt{2} }{M_D^{\frac{D-2}{2}}} g_{a_1 a_2 i_3} \left( 2  \epsilon_1 \ccdot \epsilon_2 \, \epsilon_1 \ccdot \epsilon_3 \, \epsilon_2 \ccdot p_1 -  (\epsilon_1 \ccdot \epsilon_2)^2  \, \epsilon_3\ccdot p_1 -(1 \leftrightarrow 2) \right) \label{eq:hhA} \, ,
\ee
where we have rescaled $A^{\mu}_i \rightarrow A^{\mu}_i /\sqrt{2}$ to canonically normalize the vector kinetic terms.
This vertex vanishes if $a_1 = a_2$ since $\lambda_{a_1 a_2 i_3}$ is antisymmetric in its first two indices, so the two massive spin-2 particles must be distinct for this interaction to be nonvanishing. The vector $A^{\mu}_{i_3}$ is massless when $i_3 \in I_{\rm{Killing}}$, so in this case the on-shell vertex should also be gauge invariant. The on-shell gauge variation is given by
\be
\delta \mathcal{V}(1_h^{a_1}, \, 2_h^{a_2}, \, 3_A^{i_3})\Big|_{\rm on-shell} =\frac{\sqrt{2} }{M_D^{\frac{D-2}{2}}}  \left( \lambda_{a_2} - \lambda_{a_1} \right) g_{a_1 a_2 i_3}  (\epsilon_1 \ccdot \epsilon_2)^2.
\ee
We can show that this indeed vanishes for Killing vectors using the identity
\be
 0 = \intN \partial^n \psi_{a_1} \partial^m \psi_{a_2} \left( \nabla_{n} Y_{m, i_3 }+\nabla_{m} Y_{n, i_3}\right) =  \left( \lambda_{a_2} - \lambda_{a_1} \right) g_{a_1 a_2 i_3} \, , \quad i_3 \in I_{\rm{Killing}},
\ee
where the first equality follows from Killing's equation and the second equality follows after integrating by parts. This shows that the two massive spin-2 particles coupling to a massless vector must have identical mass as well as $a_1 \neq a_2$, so such couplings require degenerate eigenvalues.

The vertices involving two massive spin-2 particles and one of the various scalars after canonically normalizing the scalar kinetic terms are
\begin{align}
\mathcal{V}(1_h^{a_1}, \, 2_h^{a_2}, \, 3_{\phi}^{a_3})  = &  \frac{-i\sqrt{D-2}\left(\lambda_{a_3}-2(\lambda_{a_1}+\lambda_{a_2})+\lambda_{a_3}^{-1}(\lambda_{a_1}-\lambda_{a_2})^2\right)}{2M_D^{\frac{D-2}{2}}\sqrt{(N-1)(d-1)}}g_{a_1 a_2 a_3} \, (\epsilon_1 \ccdot \epsilon_2)^2 \, ,\\
\mathcal{V}(1_h^{a_1}, \, 2_h^{a_2}, \, 3_\phi^0)  = & \frac{2i}{M_d^{\frac{d-2}{2}}} \sqrt{\frac{D-2}{N(d-2)}} \lambda_{a_1} \delta_{a_1 a_2}\, (\epsilon_1 \ccdot \epsilon_2)^2  \, , \\
\mathcal{V}(1_h^{a_1}, \, 2_h^{a_2}, \, 3_{\phi}^{\mathcal{I}})  = & \frac{2i}{M_D^{\frac{D-2}{2}}}g_{a_1 a_2 \mathcal{I}} \, (\epsilon_1 \ccdot \epsilon_2)^2 \, .
\end{align}
Recall that we restrict to $d>2$ and that for $N=1$ there are no $\phi^a$ scalars, so the coupling constants are all well defined.

\subsubsection{Quartic interactions}

We also need the interactions between four massive spin-2 fields. From the higher-dimensional quartic interactions in Eq.~\eqref{eq:GR4} we get
\begin{align}
\mathcal{L}_{hhhh} =&\frac{1}{M_D^{D-2}} \sum_{a_1, a_2, a_3, a_4} \Big[ g_{a_1 a_2 a_3 a_4}\Big(  h^{a_1 \mu \nu} h^{a_2 \lambda \rho}\nabla_{\mu} h^{a_3}_{\lambda \sigma} \nabla_{\rho} h^{a_4}_{\nu}{}^{\sigma}+h^{a_1 \mu \sigma} h^{a_2}_{\sigma}{}^{\nu} \big[ 2\nabla^{\lambda} h^{a_3}_{\mu}{}^{\rho} \nabla_{[\rho} h^{a_4}_{\lambda]}{}^{\nu} \nn \\ 
                  \qquad & +2 \nabla^{\lambda} h^{a_3}_{\mu \rho}  \nabla_{\nu} h^{a_4}_{\lambda}{}^{\rho}-  \nabla_{\mu} h^{a_3}_{\lambda \rho} \nabla_{\nu} h^{a_4 \lambda \rho} \big]  -\lambda_{a_1} h^{a_1}_{\mu}{}^{\lambda} h^{a_2}_{\nu \lambda} h^{a_3}_{\rho}{}^{\mu} h^{a_4 \rho \nu}\Big) \nn \\
&+\sum_a \frac{\lambda_a}{2} g_{a_1 a_2 a} g_{a_3 a_4 a} h^{a_1}_{\mu}{}^{\lambda} h^{a_2}_{\nu \lambda} h^{a_3}_{\rho}{}^{\mu} h^{a_4 \rho \nu}\Big]+ \ldots \, . 
\end{align}
The corresponding four-point  vertex is given by
\begin{align}
\mathcal{V}(1_h^{a_1}, \, 2_h^{a_2}, \, 3_{h}^{a_3}, \, 4_h^{a_4})  = &\frac{i}{M_D^{D-2}} \Big[ g_{a_1 a_2 a_3 a_4} \epsilon_1 \ccdot \epsilon_2 \,  \epsilon_3 \ccdot \epsilon_4 \Big( ( p_1 \ccdot p_2 -\lambda_{a_1}) \, \epsilon_1 \ccdot \epsilon_4 \, \epsilon_2 \ccdot \epsilon_3 \nn \\
& + \epsilon_1\ccdot  p_4  \left[ \epsilon_3 \ccdot \epsilon_4 \, \epsilon_2 \ccdot p_3-  \epsilon_2 \ccdot \epsilon_4 \,  \epsilon_3 \ccdot p_2- \epsilon_2 \ccdot \epsilon_3( \epsilon_4 \ccdot p_1 +2  \epsilon_4 \ccdot p_3)  \right]\Big) \nn \\
& + \frac{1}{2 } \lambda_a g_{a_1 a_2 a} g_{a_3 a_4 a} \epsilon_1 \ccdot \epsilon_2 \,  \epsilon_3 \ccdot \epsilon_4 \,  \epsilon_1 \ccdot \epsilon_4 \, \epsilon_2 \ccdot \epsilon_3 + \dots \Big] + {\rm 23 \, \, permutations} \, .
\end{align}

\subsection{Amplitudes and sum rules}

We now calculate some four-point amplitudes. In GR the four-point amplitude grows like $\sim E^2$ at high energies in any dimension. By imposing that the amplitudes of massive KK modes have this same high-energy behavior, we can derive sum rules that constrain the cubic couplings and masses.\footnote{The sum rules for one-dimensional compactifications have been studied in Ref.~\cite{Chivukula:2019zkt}.}

We restrict to amplitudes with longitudinal polarizations, since these have the worst high-energy behavior and thus lead to the most interesting constraints. The longitudinal massive spin-2 polarization tensor can be written in terms of the longitudinal spin-1 polarization vector \eqref{eq:spin1pol} as
\be \label{eq:spin2pol}
\epsilon^{\mu \nu}_{L,j} (p_j) = \sqrt{\frac{d-1}{d-2}}\left[ \epsilon^{\mu}_{L,j} \epsilon^{\nu}_{L,j} - \frac{1}{d-1}\left(\eta^{\mu \nu} + \frac{p_j^{\mu} p_j^{\nu}}{m_j^2} \right) \right].
\ee

\subsubsection{Identical four-point tensor scattering}
Consider first the scattering of identical massive spin-2 particles,
\be
h_L^{a_1} h_L^{a_1} \rightarrow h_L^{a_1} h_L^{a_1}.
\ee
This amplitude receives contributions from the exchange of the graviton, massive spin-2 particles, and the various spin-0 particles. 

The leading high-energy piece scales like $ \sim E^{10}$ and vanishes by expanding the quartic coupling in terms of the cubic couplings as in Eq.~\eqref{eq:assoc}, just as for the leading $ \sim E^4$ amplitudes in the dimensionally reduced YM theories. 
The next term grows like $\sim E^8$ and is proportional to the sum rule in Eq.~\eqref{eq:sumrule1b} that we found from the $\sim E^2$ part of the YM amplitudes. So the would-be leading and subleading high-energy terms in GR just reproduce the sum rules found from YM. 

At order $E^6$ we get something new compared to YM. There is one additional sum rule that must be satisfied for the amplitude to vanish, which for $N>1$ can be written as
\be \label{eq:grE6}
\sum_a \left((4-3N)N \lambda_a^2+4(N^2-3) \lambda_{a_1} \lambda_a +16\lambda_{a_1}^2\right) g_{a_1 a_1 a}^2 +16N(N-1) \sum_{\mathcal{I}} g_{a_1 a_1 \mathcal{I}}^2 =0.
\ee
To get the sum rule in this simple form we have freely added multiples of the earlier sum rules, which is why the there is no $d$ dependence nor constant contributions from the zero modes. The first term represents the remaining contributions from the exchange of massive tensors and the massive scalars $\phi^a$, while the second term corresponds to the contribution from exchanging the scalars $\phi^{\mathcal{I}}$. When $N=1$ we instead get the sum rule
\be
\sum_a \left(4 \lambda_{a_1} -\lambda_a\right) \left(4 d \lambda_{a_1} +(2-3d)\lambda_a \right) g^2_{a_1 a_1 a} =0.
\ee
It can be straightforwardly checked that this is satisfied for the circle, for any value of $d$, following the discussion in Section~\ref{sec:circle}.

At order $E^4$ there is another new sum rule that for $N>1$ can be written as
\be \label{eq:grE4}
\sum_a \lambda_a (\lambda_a-4 \lambda_{a_1})\left((3N-2)\lambda_a-4N \lambda_{a_1} \right)g_{a_1 a_1 a}^2 +16(N-1) \sum_{\mathcal{I}} \lambda_{\mathcal{I}}g_{a_1 a_1 \mathcal{I}}^2 =0,
\ee
where we have again simplified by adding multiples of the earlier sum rules. For $N=1$ there are additional sum rules whose explicit form is unenlightening but they are included in the ancillary file and can be easily verified for the circle.

These sum rules must be satisfied for the massive spin-2 amplitudes to grow no faster than $\sim E^2$ at high energies. Below we prove that these are indeed satisfied on any closed Ricci-flat manifold. That we have additional sum rules compared to YM can be understood from the fact that GR is defined on a more restrictive class of background spacetimes. 

We now discuss some consequences of these sum rules. Consider first Eq.~\eqref{eq:grE6}. For this to be satisfied in combination with Eq.~\eqref{eq:sumrule1b}, there must be a term in the first sum that is non-positive with a nonvanishing cubic coupling, which implies that for each $a_1$ there exists some $a^*$ such that
\be \label{eq:grE6implies}
g_{a_1 a_1 a^*} \neq 0 \quad {\rm and } \quad  \frac{8}{3-N^2+(N-1)\sqrt{N^2+2N+9}} \lambda_{a_1} \leq \lambda_{a^*}.
\ee
This reduces to the earlier bound in Eq.~\eqref{eq:YMbound} as $N \rightarrow \infty$, which is a good consistency check, but for finite $N$ it gives a stronger condition for closed Ricci-flat manifolds. For example, for $N=2$ we get that for every $a_1$ there exists some $a^*$ such that
\be
g_{a_1 a_1 a^*} \neq 0 \quad {\rm and } \quad  \lambda_{a^*} \geq \frac{1}{2}(1+\sqrt{17}) \lambda_{a_1} \approx 2.56 \lambda_{a_1}.
\ee

Now consider Eq.~\eqref{eq:grE4}. Combined with Eq.~\eqref{eq:sumrule1b} it implies that for stable compactifications, i.e. for internal manifolds with $\lambda_{\cal I}\geq 0$, which is all known compact Ricci-flat manifolds, there must exist an exchanged massive spin-2 particle with mass below some upper bound.\footnote{This is the geometric analogue of finding an upper bound $\Delta_{\rm max}$ on the dimension of the first nontrivial scalar operator appearing in the $\phi \times \phi$ OPE of a given scalar operator $\phi$ in a CFT~\cite{Poland:2018epd}.} Indeed, if $\lambda_{\mathcal{I}} \geq 0$ then the first sum must contain a non-positive term, so for each $a_1$ there exists some $a^*$ such that
\be \label{eq:grE4implies}
g_{a_1 a_1 a^*} \neq 0 \quad {\rm and } \quad   \frac{4N}{3N-2} \lambda_{a_1} \leq \lambda_{a^*} \leq 4\lambda_{a_1} .
\ee
The eigenvalue expansion of the square of a real scalar eigenfunction thus takes the form
\be
\psi_{a_1}^2 = \frac{1}{V} +  g_{a_1 a_1 a^*} \psi_{a^*}+ \dots \, , \quad \lambda_{a^*} \leq 4 \lambda_{a_1}.
\ee

A corollary of \eqref{eq:grE4implies} is that the ratios of consecutive nonzero eigenvalues of the scalar Laplacian on a closed Ricci-flat manifold with $\lambda_{\mathcal{I}} \geq0$ are bounded above by four,
\be \label{eq:eigenbound}
\frac{\lambda_{k+1}}{\lambda_k} \leq 4,
\ee
where $\lambda_k$ is the $k$\textsuperscript{th} nonzero eigenvalue.
This bound  is new as far as we can tell and applies to all known closed Ricci-flat manifolds, since they all have covers that admit parallel spinors and thus have positive-definite Lichnerowicz operators \cite{Wang91, Gibbons:2002th, Dai2005}.
It is saturated  in every dimension by the first distinct nonzero eigenvalues on certain tori, as we discuss below, so it is the optimal bound of this form. It is similar in form to the extended Payne--P\'olya--Weinberger conjecture (reviewed in Appendix~\ref{sec:deVerdiere}), which gives an upper bound on the ratio of consecutive eigenvalues of the scalar Laplacian for the Dirichlet problem on bounded domains in $\mathbb{R}^N$. 
By de Verdi\`ere's theorem \cite{deVerdiere1986c, deVerdiere1987a}, reviewed in Appendix~\ref{sec:deVerdiere}, such a bound is not possible for general closed manifolds, i.e. closed manifolds that are not necessarily Ricci-flat.

Phrased in terms of the particle spectrum, Eq.~\eqref{eq:eigenbound} says that adjacent massive spin-2 particles have masses that differ by at most a factor of two. This provides a sharp bound on the possible spectra of KK excitations of the graviton.  It means that there can be no large gaps between the massive modes (relative to their masses), so one cannot integrate out all but a finite number of massive spin-2 particles from the spectrum to obtain an effective multi-metric field theory with a finite number of spin-2 modes and a strong coupling scale that is parametrically larger than the masses, such as  the theories of Refs.~\cite{Hassan:2011zd,Hinterbichler:2012cn}.
Although we started with higher-dimensional GR, this statement follows purely from geometry and so applies to the spectrum of more general compactifications of gravity. In particular, it applies to the massive excitations of the graviton in smooth Calabi--Yau compactifications of string theory and in $G_2$ compactifications of M-theory in the low-energy limit.

\subsubsection{Proving the sum rules}
As with YM, we can prove all of the sum rules directly using eigenvalue and Hodge/tensor decompositions to derive associativity relations. In particular, Eq.~\eqref{eq:grE6} can be obtained from the relation
\be \label{eq:grE6identity}
\int_{\mathcal{N}} \contraction{}{\partial_m\psi_{a_1}}{}{\partial_n\psi_{a_1}}
\contraction{\partial_m\psi_{a_1} \partial_n\psi_{a_1}}{\partial^m \psi_{a_1}}{}{\partial^n \psi_{a_1}}
\partial_m\psi_{a_1}\partial_n\psi_{a_1}\partial^m \psi_{a_1}\partial^n \psi_{a_1}
=
\int_{\mathcal{N}} \contraction{}{\partial_m\psi_{a_1}}{\partial_n\psi_{a_1}}{\partial^m \psi_{a_1}}
\contraction[2ex]{\partial_m\psi_{a_1}}{\partial_n\psi_{a_1}}{\partial^m \psi_{a_1}}{\partial^n \psi_{a_1}}
\partial_m\psi_{a_1}\partial_n\psi_{a_1}\partial^m \psi_{a_1}\partial^n \psi_{a_1} \, .
\ee
To evaluate the left-hand side we need the following decomposition:
\begin{align}
& \frac{1}{2}\left(\partial_m \psi_{a_1} \partial_n \psi_{a_2}+\partial_m \psi_{a_2} \partial_n \psi_{a_1} \right) = \sum_{\mathcal{I}}g_{a_1 a_2}{}{}^{ \mathcal{I}} h^{TT}_{mn, \mathcal{I}}+ \sum_{i \notin I_{\rm Killing}}\frac{(\lambda_{a_2}-\lambda_{a_1})g_{a_1 a_2}{}{}^i}{2\left(\lambda_i - \frac{2R_{(N)}}{N}\right)} \left( \nabla_m Y_{n, i} +\nabla_n Y_{m, i}\right)\nn \\
& + \sum_{a \notin I_{\rm conf.}}\frac{\left((N-2)\lambda_a^2+2 \lambda_a (\lambda_{a_1}+\lambda_{a_2})-N (\lambda_{a_1}-\lambda_{a_2})^2 \right) g_{a_1 a_2}{}{}^a}{4 \lambda_a \left((N-1)\lambda_a-R_{(N)}\right)} \left(\nabla_{m} \nabla_n \psi_a - \frac{1}{N} \Box \psi_a \gamma_{mn} \right) \nn \\
& + \frac{1}{2N} \gamma_{mn} \sum_a ( \lambda_{a_1}+\lambda_{a_2} -\lambda_a )g_{a_1 a_2}{}{}^a \psi_a  +\frac{\lambda_{a_1}}{NV} \gamma_{mn} \delta_{a_1 a_2} \, ,
\end{align}
which we have written for a general closed Einstein manifold with Ricci scalar $R_{(N)}$.
Using this to evaluate \eqref{eq:grE6identity} and using the earlier sum rule \eqref{eq:sumrule1b} to simplify, we obtain the sum rule
\begin{align}
&\sum_a \left((4-3N)N \lambda_a^2+4(N^2-3) \lambda_{a_1} \lambda_a +16\lambda_{a_1}^2\right) g_{a_1 a_1 a}^2 +16N(N-1) \sum_{\mathcal{I}} g_{a_1 a_1 \mathcal{I}}^2 \nn \\
&+\sum_{a \notin I_{\rm conf.}} \frac{R_{(N)} \left(4 \lambda_{a_1}+(N-2)\lambda_a\right)^2 g_{a_1 a_1 a}^2 }{ (N-1)\lambda_a- R_{(N)}} =0,
\end{align}
which reduces to Eq.~\eqref{eq:grE6} when $R_{(N)}=0$. 
When $R_{(N)}>0$, the additional term on the final line is positive by the Lichnerowicz bound \eqref{eq:lich2}, so the bound in  Eq.~\eqref{eq:grE6implies} holds with a strict inequality for Einstein manifolds with positive Ricci curvature.

Similarly, we can derive \eqref{eq:grE4} from the following associativity identity:
\be \label{eq:grE4identity}
\int_{\mathcal{N}} \contraction{}{\partial_m\psi_{a_1}}{}{\partial_n\psi_{a_1}}
\contraction{\partial_m\psi_{a_1} \partial_n\psi_{a_1}}{\Delta_L \big( \partial^m \psi_{a_1}}{}{\partial^n \psi_{a_1}\big)}
\partial_m\psi_{a_1}\partial_n\psi_{a_1}\Delta_L \left( \partial^m \psi_{a_1}\partial^n \psi_{a_1} \right)
=
\int_{\mathcal{N}} \contraction{}{\partial_m\psi_{a_1}}{\partial_n\psi_{a_1}}{\Delta_L \big(\partial^m \psi_{a_1}}
\contraction[2ex]{\partial_m\psi_{a_1}}{\partial_n\psi_{a_1}}{ \Delta_L \big(\partial^m \psi_{a_1}}{\partial^n \psi_{a_1} \big)}
\partial_m\psi_{a_1}\partial_n\psi_{a_1} \Delta_L \left( \partial^m \psi_{a_1}\partial^n \psi_{a_1} \right) ,
\ee
where we evaluate the right-hand side by first expanding the Lichnerowicz Laplacian using
\be
 \Delta_L \left( \partial_m \psi_{a_1}\partial_n \psi_{a_2} \right) = \left(\lambda_{a_1}+\lambda_{a_2}\right) \partial_m \psi_{a_1}\partial_n \psi_{a_2}- 2  \nabla_m \nabla^p \psi_{a_1}\nabla_n \nabla_p \psi_{a_2}-2R_{m p n q}\partial^p \psi_{a_1} \partial^q \psi_{a_2},
\ee
which holds on general Einstein manifolds.
This gives a generalization of \eqref{eq:grE4} that is also valid for Einstein manifolds with nonzero $R_{(N)}$,
\begin{align}
& \sum_a \lambda_a (\lambda_a-4 \lambda_{a_1})\left((3N-2)\lambda_a-4N \lambda_{a_1} \right)g_{a_1 a_1 a}^2 +16(N-1) \sum_{\mathcal{I}} \lambda_{\mathcal{I}}g_{a_1 a_1 \mathcal{I}}^2 \nn \\
&+\sum_{a \notin I_{\rm conf.}} \frac{R_{(N)} \lambda_a \left(4 \lambda_{a_1}+(N-2)\lambda_a\right)^2 g_{a_1 a_1 a}^2 }{N\left( (N-1)\lambda_a- R_{(N)}\right)}=0.
\end{align}
The correction term is again positive for $R_{(N)}> 0$, so Eqs.~\eqref{eq:grE4implies} and \eqref{eq:eigenbound} hold with strict inequalities for closed Einstein manifolds with positive curvature and a non-negative Lichnerowicz operator. If a positively curved Einstein space admits a nonzero Killing spinor then there is a lower bound on the Lichnerowicz eigenvalues  \cite{Wang91, Gibbons:2002th},
\be
\lambda_{\mathcal{I}} \geq \frac{R_{(N)}}{4N(N-1)} \left( 16-(5-N)^2\right).
\ee 
The Lichnerowicz spectrum is thus nonnegative on any positively curved Einstein space with dimension $N \leq 9$ that admits a nonzero Killing spinor. This implies, for example, that the eigenvalues of the scalar Laplacian on compact smooth Sasaki--Einstein manifolds with $N \leq 9$ satisfy the bound $\lambda_{k+1} < 4 \lambda_k$.

\subsubsection{General four-point tensor scattering}
Now consider the general four-point scattering of longitudinally-polarized massive spin-2 modes,
\be \label{eq:h1h2h3h4}
h_L^{a_1} h_L^{a_2} \rightarrow h_L^{a_3} h_L^{a_4}. 
\ee
At order $E^{10}$, the sum rules follow from the zero-derivative associativity relations in Eq.~\eqref{eq:OPEintegral}.
The $\sim E^8$ term then gives multiple two-derivative sum rules, including the scattering sum rule in Eq.~\eqref{eq:sumrule1} that was found from YM.
At orders $E^6$ and $E^4$, we find many additional sum rules that generalize those found above for equal mass scattering, but they are much more complicated and so we relegate them to an ancillary file. We have checked that they can all be derived using the various eigenfunction and Hodge decompositions to evaluate integrals of total derivatives. This proves them directly and verifies that the amplitude grows no faster than $\sim E^2$. 
In the ancillary file we provide the sum rules plus the full dimensionally reduced GR amplitude with external massive gravitons.

\subsubsection{Strong coupling scale}

Since both the higher- and lower-dimensional theories have amplitudes that grow with energy, they are still effective field theories that become strongly coupled at some UV energy scale.  The UV strong coupling scale of the dimensionally reduced theory must be the same as that of the higher-dimensional theory, since the two theories are equivalent.   As we have seen, the amplitudes in the dimensionally reduced theory grow as $\sim E^2$ to match the energy scaling of the higher-dimensional theory,
\be\label{aeequigse}
\mathcal{A}(E) \sim \frac{E^2}{M_d^{d-2}},
\ee 
where  $M_d$ is the $d$-dimensional Planck scale.\footnote{As in the YM case, we assume for this schematic argument that various dimensionless combinations of geometrical quantities are not parametrically small.}
The lower-dimensional theory thus seems to become strongly coupled around the scale $M_d$.   However, this cannot be correct since the dimensionally reduced theory is equivalent to the higher-dimensional theory, which becomes strongly coupled at a much lower scale given by the $D$-dimensional Planck mass, $M_D$.
The reason that $M_D\ll M_d$ is due to the relation $M_d^{d-2} = V M_D^{D-2}$ and the fact that the volume of the internal manifold should satisfy ${ V^{-1/N}}\ll M_D$ for the compactification to make sense in the effective field theory.

The way to see this lower cutoff in the lower-dimensional theory is again by scattering states that are normalized superpositions of KK modes \cite{SekharChivukula:2001hz, Chivukula:2019rij}. For the four-point scattering of superposed states of $n$ modes we get the amplitude
\be 
{\cal A}^{\rm superposition}(E) \sim n {E^2  \over M_{d}^{d-2}} \label{speccute}.
\ee
Using Weyl's law \eqref{nrelaeqn} to find the maximum number of spin-2 states that can be scattered and choosing the cutoff $\Lambda$ so that the amplitude saturates the unitarity bound gives
\be
{\cal A}^{\rm superposition}(\Lambda) \sim {1\over \Lambda^{d-4}} \implies \Lambda \sim M_D,
\ee
confirming that the strong coupling scale is given by the higher-dimensional Planck scale.
As in the YM case, we see that the strong coupling scale is parametrically lower due to the large number of particles in the KK tower.

\subsection{Examples}
We now discuss some explicit examples of internal manifolds to demonstrate the additional sum rules found in this section. 

\subsubsection{Tori}
The number of independent non-constant transverse, traceless symmetric tensors on an $N$-dimensional torus $\mathbb{T}^N$ is $(N+1)(N-2)/2$.
Each of these leads to a tower of massive scalars in the dimensionally reduced theory. There are additionally $(N+2)(N-1)/2$ zero modes, corresponding to the finite number of massless shape moduli in lower dimensions.

The simplest manifold with transverse traceless tensors is thus the ordinary square flat torus, $\mathbb{T}^2$. This has two zero-mode tensors, 
\be
h_{mn, +}^{TT} =
\frac{1}{2\sqrt{2}\pi}
\begin{pmatrix} 
0 &1 \\
1 & 0 
\end{pmatrix}, \quad
h_{mn, -}^{TT} =
\frac{1}{2\sqrt{2}\pi}
\begin{pmatrix} 
1 &0 \\
0 & -1 
\end{pmatrix}.
\ee
Consider the four-point amplitude with identical external massive spin-2 particles described by the scalar eigenfunctions $\psi_{\vec{a}_1}$ with $\vec{a}_1 = (2 k_1, 2 k_2)$ and $k_i \in \mathbb{Z}_{>0}$, using the notation of Section~\ref{sec:tori1}. The triple overlap integrals $g_{\vec{a}_1 \vec{a}_1 \mathcal{I}}$ are then given by
\be
g_{\vec{a}_1 \vec{a}_1 + } = 0, \quad g_{\vec{a}_1 \vec{a}_1 -} = \frac{1}{2 \sqrt{2} \pi} (k_1^2-k_2^2), 
\ee
where $\lambda_{\pm} =0$.
From Eq.~\eqref{eq:torusdata} we also have the nonvanishing couplings $g_{\vec{a}_1 \vec{a}_1 \vec{a}_r}$ given by
\be
(\lambda_{\vec{a}_r}, g_{\vec{a}_1 \vec{a}_1 \vec{a}_r} ) \in \left\{\left(4k_1^2, \frac{1}{2 \sqrt{2} \pi} \right),\left(4k_2^2, \frac{1}{2 \sqrt{2} \pi} \right),\left(4k_1^2+4k_2^2, \frac{1}{4 \pi} \right) \right\} .
\ee
This amplitude thus involves the exchange of a massless scalar and three massive spin-2 particles with squared masses $m^2_{\vec{a}_r} = \lambda_{\vec{a}_r}$. 
Substituting these values into the sum rules \eqref{eq:grE6} and \eqref{eq:grE4} verifies that they provide a solution and thus lead to an amplitude with $\sim E^2$ high-energy behavior.

\begin{figure}[!ht]
\begin{center}
\epsfig{file=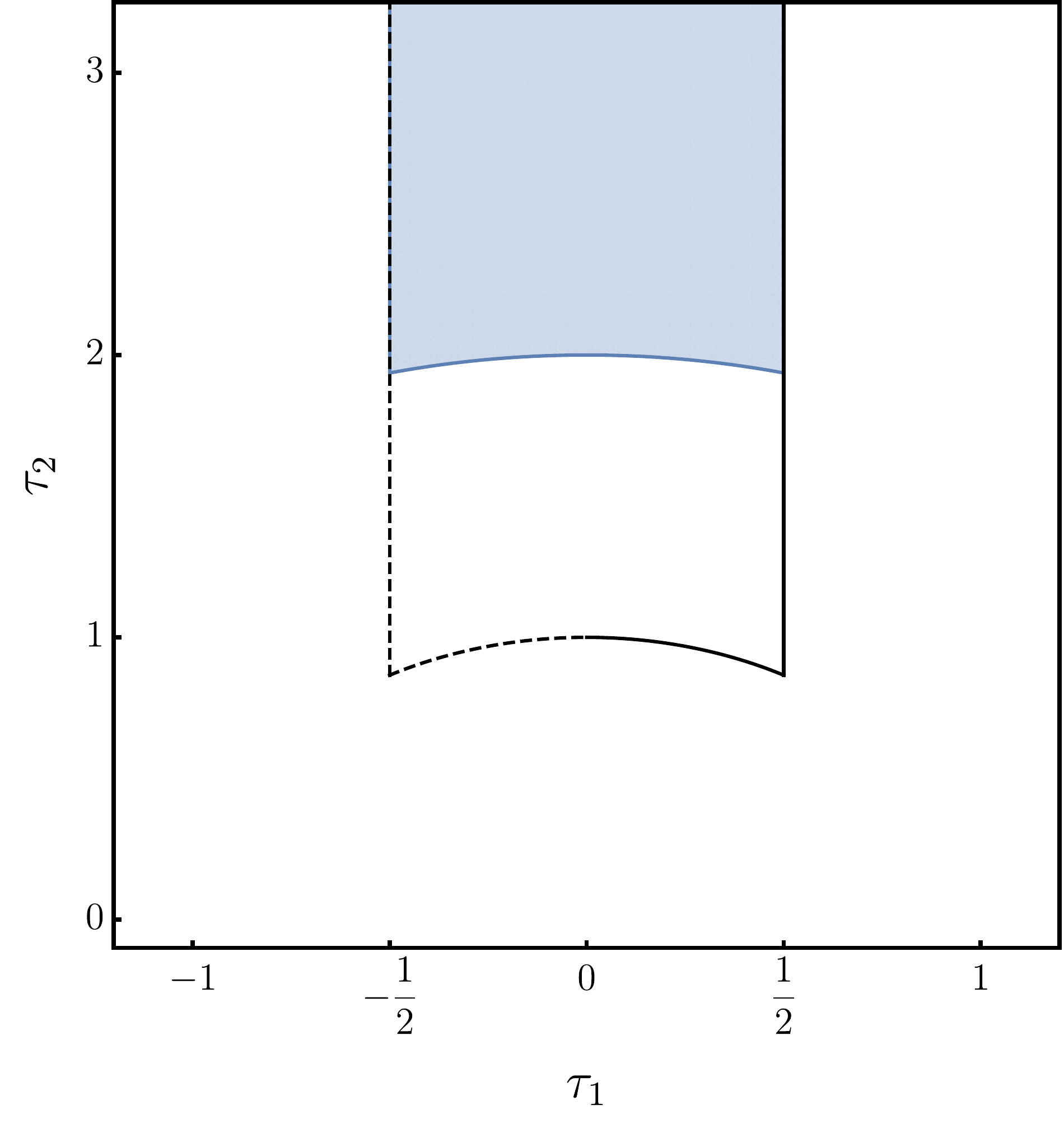,scale=.4}
\caption{\small The fundamental domain $\mathcal{F}(\tau)$ of the 2-torus with the shaded region showing tori whose scalar Laplacian eigenvalues satisfy $\lambda_2= 4 \lambda_1$.}
\label{fig:tori}
\end{center}
\end{figure}

Now we discuss the ratios of consecutive eigenvalues of the scalar Laplacian on tori.
The maximum ratio of consecutive eigenvalues on a square torus is two, but on more general tori this ratio can be as large as four.
A general 2-torus can be written as $\mathbb{R}^2/\Gamma$, where $\Gamma$ is a lattice generated by integer linear combinations of the vectors $(0,1)$ and $\tau=(\tau_1, \tau_2)$, where $\tau_i \in \mathbb{R}$. Eigenfunctions are then labeled by elements of the dual lattice. The moduli space of conformally inequivalent 2-tori can be parameterized by the following standard fundamental domain, which is shown in Figure~\ref{fig:tori}:
\be
\mathcal{F}(\tau) = \{ \tau_1, \tau_2 \in \mathbb{R}: \, 1/2<\tau_1\leq 1/2, \quad \tau_2>0, \quad |\tau|^2 \geq 1, \quad \rm{with}\quad  \tau_1\geq0 \quad {\rm if} \quad |\tau|^2 = 1  \},
\ee
where $|\tau|^2 = \tau_1^2+\tau_2^2$.
Since the ratios of eigenvalues do not change under diffeomorphisms and local Weyl transformations in two dimensions, we can restrict to a fundamental domain to find all possible such ratios. 
Of the tori in $\mathcal{F}(\tau) $, those with $|\tau|^2 \geq 4$ saturate the bound \eqref{eq:eigenbound}, as indicated by the shaded region in Figure~\ref{fig:tori}. In fact, for every $N$ there exist $N$-tori for which $\lambda_2 =4\lambda_1$. 
Consider, for example, the ``long" rectangular tori $\mathbb{R}^N/\Gamma$ with $\Gamma$ generated by the vectors $\{p \vec{e}_1, \vec{e}_2, \dots, \vec{e}_N\}$ where $\vec{e}_j$ is the unit $N$-vector in the $j$\textsuperscript{th} direction and $p \geq 2$. The first three non-zero eigenvalues of these tori are $\lambda_1=(2 \pi/ p)^{2}$, $\lambda_2 =4(2 \pi/ p)^{2}$, and $\lambda_3 =(2 \pi)^2$, so these saturate the bound \eqref{eq:eigenbound}, which shows that it is optimal in every dimension. 

\subsubsection{Spheres}

Consider the spheres $\mathbb{S}^{N}$ with the round metric and $N \geq 2$. These have positive Lichnerowicz spectrum so the bound \eqref{eq:eigenbound} applies, even though we cannot use them as the internal space for a flat compactification of GR. The possible scalar eigenvalues are given by $l(l+N-1)$ where $l \in \mathbb{Z}_{\geq 0}$, so the maximum ratio of consecutive eigenvalues is
\be
\max_{k} \left( \frac{\lambda_{k+1}}{\lambda_k} \right) = \frac{2(N+1)}{N} \leq 3,
\ee
which is consistent with \eqref{eq:eigenbound}.

\subsubsection{Fermat quintic}
There are no nontrivial Calabi--Yau manifolds for which the spectrum and eigenfunctions of the scalar Laplacian are known analytically, but the lowest ones can be determined numerically using Donaldson's algorithm to approximate the Ricci-flat K\"ahler metric \cite{Donaldson2005, Douglas:2006rr}, as shown in Ref.~\cite{Braun:2008jp}.

Consider the example of the Fermat quintic threefold, which is the submanifold of $\mathbb{C}P^4$ defined by the embedding
\be
\sum_{j=1}^5 Z_j^5=0,
\ee
where $Z_1, \dots, Z_5$ are the homogeneous coordinates on $\mathbb{C}P^4$ with the identification $Z_j \sim \lambda Z_j$ for nonzero complex  $\lambda$. With the volume normalization $V=1$,  Ref.~\cite{Braun:2008jp} found numerically that the first few distinct nonzero eigenvalues of the scalar Laplacian on this manifold
are
\be
\lambda_k  \in \{41.1\pm 0.4, 78.1 \pm 0.5, 82.1 \pm 0.3, 94.5 \pm 1, 102 \pm 1\}.
\ee
These eigenvalues easily satisfy the bound \eqref{eq:eigenbound} and this is also true for the spectra of the other Calabi--Yau manifolds studied in Ref.~\cite{Braun:2008jp}.

\subsection{Bottom-up sum rules}

We have so far assumed in this section that our starting theory is pure GR in higher dimensions. From this starting point we proceeded in a top-down manner to derive the lower-dimensional vertices and verified that the massive spin-2 amplitudes grow no faster than $\sim E^2$.  An interesting and more difficult question is whether the converse holds, i.e. is any theory of massive spin-2 particles with $\sim E^2$ amplitudes and no higher-spin particles equivalent to an extra dimensional theory of gravity? In the remainder of this section we address one small aspect of this question by performing a bottom-up construction of certain massive spin-2 amplitudes. Note that although we use the same notation for labeling particles as in the earlier part of this section, here the labels do not necessarily correspond to eigenfunctions of Laplacian operators.

\subsubsection{Interactions}
Our goal is to study theories involving fields with spin two or less that have amplitudes growing no faster than $\sim E^2$, remaining agnostic about whether or not they come from higher-dimensional gravity. For simplicity we restrict to $d=4$ and assume that the interactions are parity-even with at most two derivatives. This is not completely general but keeps the calculation tractable. 

With these assumptions, we will calculate the four-point amplitude for the scattering of a massive spin-2 particle $h^{\star}_{\mu \nu}$ with mass $m_{\star}$, 
\be \label{eq:hstar}
h^{\star} h^{\star} \rightarrow h^{\star} h^{\star}.
\ee
We assume that $h^{\star}_{\mu \nu}$ couples to a massless graviton, $h^0_{\mu \nu}$, a collection of massive spin-2 particles $h_{\mu \nu}^a$ with masses $m_a$ (not including $h^{\star}_{\mu \nu}$), a collection of  massive vectors $A^{i}_{\mu}$ with masses $m_i$, and a collection of scalars $\phi^{\mathcal{I}}$ with masses $m_{\mathcal{I}}$. For continuity we have labeled the particles using the same indices as in the top-down KK construction, even though these labels may now have nothing to do with geometry.\footnote{The comparison to the KK case is imperfect since there we also had scalars labeled by 0 and $a$.} The masses are assumed to be real so that the theory is stable.

The general two-derivative parity-even on-shell cubic vertices involving these fields and two or more $h_{\mu \nu}^{\star}$ legs are
\begin{align}
\mathcal{V}(1_h^{\star}, \, 2_h^{\star}, \, 3_h^{\star}) &=i a_1 m_{\star}^2 \, \epsilon_1 \ccdot \epsilon_2 \, \epsilon_1  \ccdot \epsilon_3 \, \epsilon_2 \ccdot \epsilon_3 +i a_2 \left( (\epsilon_2 \ccdot \epsilon_3)^2 (\epsilon_1 \ccdot p_2)^2+ (\epsilon_1 \ccdot \epsilon_3)^2 (\epsilon_2 \ccdot p_3)^2+ (\epsilon_1 \ccdot \epsilon_2)^2 (\epsilon_3 \ccdot p_1)^2\right)  \nn \\ 
& +i a_3 \big(  \epsilon_1 \ccdot \epsilon_3 \, \epsilon_2 \ccdot \epsilon_3 \, \epsilon_1 \ccdot p_2 \, \epsilon_2 \ccdot p_3+ \epsilon_1 \ccdot \epsilon_2 \, \epsilon_2 \ccdot \epsilon_3 \, \epsilon_1 \ccdot p_2 \, \epsilon_3 \ccdot p_1+ \epsilon_1 \ccdot \epsilon_2 \, \epsilon_1 \ccdot \epsilon_3 \, \epsilon_2 \ccdot p_3 \, \epsilon_3 \ccdot p_1 \big) ,  \\
\mathcal{V}(1_h^{\star}, \, 2_h^{\star}, \, 3_h^0) & = i b_1 (\epsilon_1 \ccdot \epsilon_2)^2 (\epsilon_3 \ccdot p_1)^2 +i b_2 \, \epsilon_1 \ccdot \epsilon_2 \, \epsilon_3 \ccdot p_1 \big(\epsilon_2 \ccdot \epsilon_3 \, \epsilon_1 \ccdot p_2+\epsilon_1  \ccdot \epsilon_3 \, \epsilon_2 \ccdot p_3 \big) \nn \\
&+i b_3 \big( \epsilon_2 \ccdot \epsilon_3 \, \epsilon_1 \ccdot p_2+\epsilon_1 \ccdot \epsilon_3 \, \epsilon_2 \ccdot p_3 \big)^2  , \\
\mathcal{V}(1_h^{\star}, \, 2_h^{\star}, \, 3_{\phi}^{\mathcal{I}}) &  = i m^2_{\star} c_{1,\mathcal{I}} \, (\epsilon_1 \ccdot \epsilon_2)^2 +ic_{2,\mathcal{I}} \, \epsilon_1 \ccdot \epsilon_2 \, \epsilon_1 \ccdot p_2 \, \epsilon_2 \ccdot p_3,  \\
\mathcal{V}(1_h^{\star}, \, 2_h^{\star}, \, 3_A^i) & =m_{\star} d_{1,i} \, \epsilon_1 \ccdot \epsilon_2 \big( \epsilon_2 \ccdot \epsilon_3 \, \epsilon_1 \ccdot p_2-\epsilon_1  \ccdot \epsilon_3 \, \epsilon_2 \ccdot p_3 \big),\\ 
\mathcal{V}(1_h^{\star}, \, 2_h^{\star}, \, 3_h^a) &=i m_{\star}^2 e_{1,a} \, \epsilon_1 \ccdot \epsilon_2 \, \epsilon_1  \ccdot \epsilon_3 \, \epsilon_2 \ccdot \epsilon_3 +i e_{2,a}\,  (\epsilon_1 \ccdot \epsilon_2)^2 (\epsilon_3 \ccdot p_1)^2 \nn \\&+i e_{3, a} \, \big( \epsilon_1 \ccdot \epsilon_2 \, \epsilon_2 \ccdot \epsilon_3 \, \epsilon_1 \ccdot p_2 \, \epsilon_3 \ccdot p_1+ \epsilon_1 \ccdot \epsilon_2 \, \epsilon_1 \ccdot \epsilon_3 \, \epsilon_2 \ccdot p_3 \, \epsilon_3 \ccdot p_1 \big) +i e_{4,a} \, \left( (\epsilon_2 \ccdot \epsilon_3)^2 (\epsilon_1 \ccdot p_2)^2+ (\epsilon_1 \ccdot \epsilon_3)^2 (\epsilon_2 \ccdot p_3)^2\right)  \nn \\ 
&+ i e_{5,a} \, \epsilon_1 \ccdot \epsilon_3 \, \epsilon_2 \ccdot \epsilon_3 \, \epsilon_1 \ccdot p_2 \, \epsilon_2 \ccdot p_3, 
\end{align}
where $a_n$, $b_n$, $c_{n, \mathcal{I}}$, $d_{n, i}$, and $e_{n, a}$ are real dimensionful coupling constants. Gauge invariance additionally gives
\be \label{eq:mincoupling}
2 b_1=b_2 = \frac{4}{M_4},
\ee
where $M_4$ denotes the four-dimensional Planck mass.

For the quartic contact interaction we construct a general ansatz consisting of polynomials in the contractions $\epsilon_i \cdot \epsilon_j$, $\epsilon_i \cdot p_j$, and $p_i \cdot p_j$ containing up to six derivatives, since this is the maximum number of derivatives that can be generated from field redefinitions at quartic order in a two-derivative theory. The polynomials must also be homogeneous in each $\epsilon_i$ with weight two. Altogether there are 95 independent polynomials satisfying these constraints. 

\subsubsection{Amplitudes and sum rules}
Putting together these ingredients, we can calculate the four-point amplitude \eqref{eq:hstar} with general polarizations and impose that it grows no worse than $\sim E^2$ at high energies. After eliminating the contact term parameters from the resulting constraints, we get a collection of sum rules involving the cubic coupling constants and masses. Using the fact that the coupling constants are real in a unitarity theory, we find that these sum rules imply the following conditions:
\begin{align} \label{eq:cons1}
a_3 & = 2 a_2, \quad b_1 =b_3, \quad 2 e_{2,a} = e_{3,a} = 2e_{4,a}= e_{5,a}, \quad a_1 = e_{1,a} = c_{2, \mathcal{I}} =d_{1,i} =0,
\end{align}
which imply that the spin-2 vertices all take the Einstein--Hilbert form. We have already seen that this follows automatically when the higher-dimensional theory is GR, but here we see that it also follows from perturbative unitarity plus our other assumptions.

After enforcing the conditions \eqref{eq:cons1}, we are left with the following three sum rules:
\begin{subequations} \label{eq:sumrules}
\begin{align}
&a_2^2 +4 b_1^2 + \frac{1}{4} \sum_{a}  \left(4-\frac{3 m_{a}^2}{m_{\star}^2} \right) e_{5,a}^2=0, \\
&a_2^2+4 b_1^2 -24  \sum_{\mathcal{I}} c_{1, \mathcal{I}}^2+\frac{1}{4} \sum_{a}  \left(\frac{5m_{a}^2}{m_{\star}^2}-4 \right) \frac{m_a^2}{m_{\star}^2} e_{5,a}^2=0, \\
& 6 \sum_{\mathcal{I}} m_{ \mathcal{I}}^2 c_{1,\mathcal{I}}^2 + \frac{1}{4} \sum_{a} \left(\frac{m_{a}^2}{m_{\star}^2}-4\right)\left(\frac{m_{a}^2}{m_{\star}^2}-1\right) m_a^2 e_{5,a}^2 =0.
\end{align}
\end{subequations}
These reduce to the sum rules found earlier in the special case when the lower-dimensional theory comes from higher-dimensional pure GR, but these are more general. 

From these sum rules we deduce that $h^{\star}_{\mu \nu}$ must couple to some $h^{a^*}_{\mu \nu}$ with $3m_{a^*}^2>4m^2_{\star}$ and to some (possibly different) $h^{a^{*}}_{\mu \nu}$ with $m_{a^{*}}^2\leq 4 m^2_{\star}$. Since $h_{\mu \nu}^{\star}$ was arbitrary to begin with, this tells us that there must be an infinite number of massive spin-2 particles and that the ratio of masses of consecutive particles is at most two.
This theory thus looks suspiciously like a theory of gravity with extra compact spatial dimensions, but we cannot rule out the existence of other kinds of solutions from this analysis. 

One consequence of these bottom-up sum rules is that the gaps between KK excitations of the graviton cannot be large compared to their masses in any compactification of gravity with a stable four-dimensional Minkowski vacuum and with no higher-derivative interactions or higher-spin particles.  This includes the Ricci-flat compactifications that we studied above as well as more general compactifications, such as compactifications with fluxes, warped products, singularities or branes.  Examples  are Freund-Rubin compactifications on $\mathbb{S}^2$ and $\mathbb{S}^3$ with the flux tuned so that lower-dimensional Minkowski space is a solution and the  Randall--Sundrum model considered in Ref.~\cite{Chivukula:2019zkt}.

\section{Summary and conclusions}
\label{sec:conclusion}

We have studied the constraints imposed by perturbative unitarity on the spectrum and cubic couplings in generic dimensional reductions of Yang--Mills theory and general relativity on closed manifolds. While the corresponding sum rules are automatically satisfied by any valid geometry, they elucidate precisely how geometries achieve the highly nontrivial cancellations required to soften the high-energy behavior of massive spin-1 and spin-2 amplitudes. We have shown how to prove these sum rules generally and have demonstrated how they work in some concrete examples.

\begin{figure}[ht]
\begin{center}
\epsfig{file=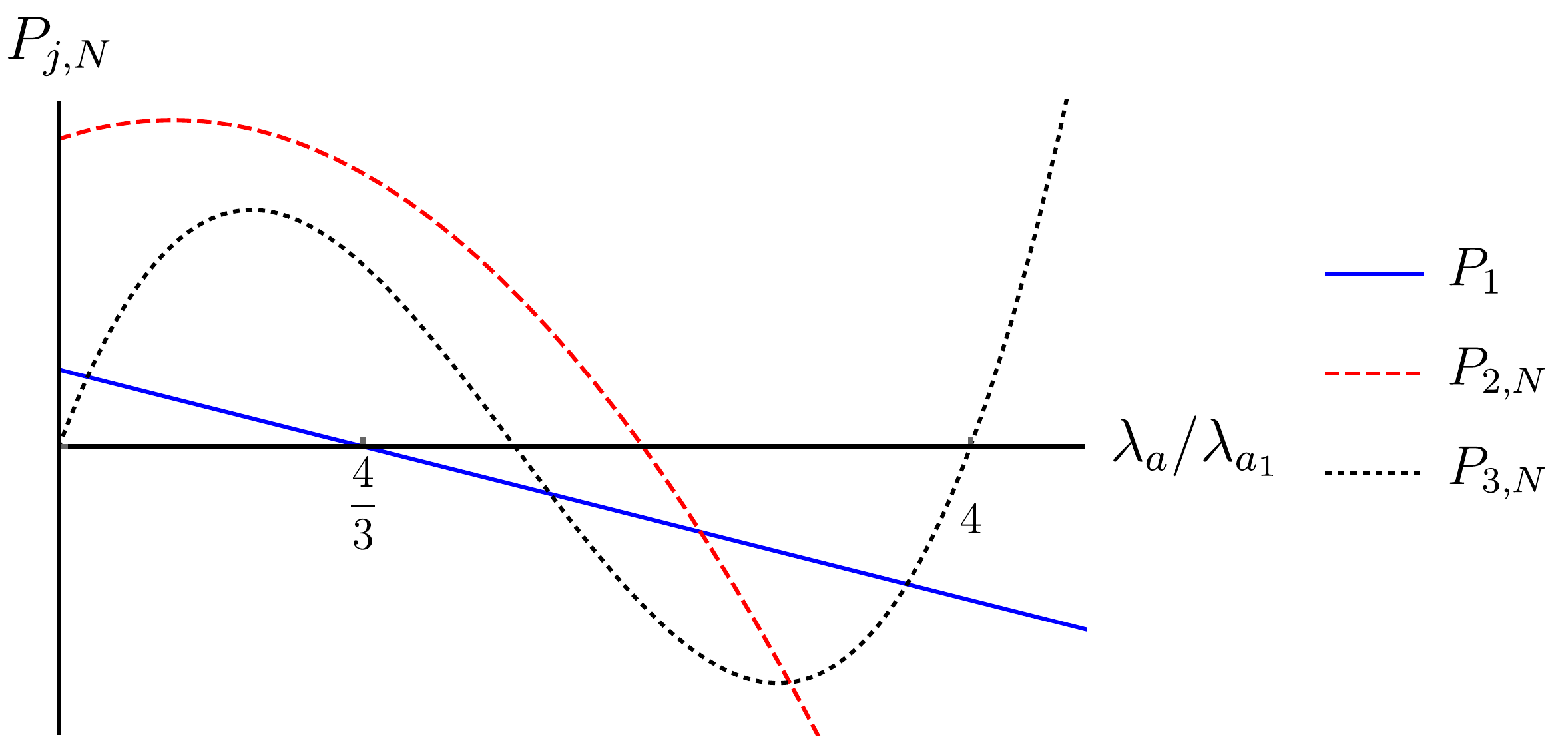,scale=.57}
\caption{\small Schematic plot of the polynomials $P_{j, N}$ appearing in the equal-mass sum rules. For each curve there is some $a^*$ such that $g_{a_1 a_1 a^*} \neq0$ and $P_{j,N}(\lambda_{a^*}/\lambda_{a_1})\leq 0$. The intersections of the nonlinear curves with the horizontal axis near the middle of the plot depend on the dimension $N$ of the internal manifold and tend to $4/3$ as $N \rightarrow \infty$.}
\label{fig:sum-rule}
\end{center}
\end{figure}

The simplest examples of the sum rules are obtained from the scattering of identical KK excitations with KK index $a_1$ and are given by
\begin{align}
\sum_a \left(4\lambda_{a_1}-3 \lambda_{a} \right)g_{a_1 a_1 a}^2+4V^{-1}\lambda_{a_1} & =0,\\
\sum_a \left((4-3N)N \lambda_a^2+4(N^2-3) \lambda_{a_1} \lambda_a +16\lambda_{a_1}^2\right) g_{a_1 a_1 a}^2 +16N(N-1) \sum_{\mathcal{I}} g_{a_1 a_1 \mathcal{I}}^2 & =0, \\
\sum_a \lambda_a (\lambda_a-4 \lambda_{a_1})\left((3N-2)\lambda_a-4N \lambda_{a_1} \right)g_{a_1 a_1 a}^2 +16(N-1) \sum_{\mathcal{I}} \lambda_{\mathcal{I}}g_{a_1 a_1 \mathcal{I}}^2 & =0,
\end{align}
where  $N$ is the dimension of the internal manifold, $V$ is its volume, $\lambda_{\bullet}$ are Laplacian eigenvalues, and $g_{\bullet \bullet \bullet}$ are triple overlap integrals of Laplacian eigenfunctions, as defined in the main text. The first constraint comes from both YM and GR amplitudes and holds for all closed manifolds, whereas the last two constraints come from GR amplitudes and hold only for closed Ricci-flat manifolds. Assuming that the eigenvalues of the Lichnerowicz operator are nonnegative, $\lambda_{\mathcal{I}} \geq0$, which is true for all known closed Ricci-flat manifolds, these can be put in the form
\be \label{eq:Pform}
\sum_a P_{1}(\lambda_a/\lambda_{a_1}) g_{a_1 a_1 a}^2 < 0, \quad \sum_a P_{j, N}(\lambda_a/\lambda_{a_1}) g_{a_1 a_1 a}^2 \leq 0, \quad j=1, 2, \quad N=1, 2, 3, \dots ,
\ee
where $P_1$ is linear and $P_{j, N}$ are order-$j$ polynomials. We schematically plot these polynomials in Figure~\ref{fig:sum-rule}. Together these sum rules imply that for each $a_1$ there is some $a^*$ such that $g_{a_1 a_1 a^*}$ is nonzero and $P_{j}(\lambda_{a^*}/\lambda_{a_1}) \leq 0$. These conditions give nontrivial constraints on the masses and interactions of KK fields that can be qualitatively understood from the plot in Figure~\ref{fig:sum-rule}. For example, from $P_{3, N}$ we deduce that every massive excitation of the graviton with mass $m$ must couple to a heavier excitation with mass less than or equal to $2m$. This gives a sharp bound on the allowed gaps in the KK spectrum of the graviton, which also applies to compactifications of string theory on Calabi--Yau manifolds, and rules out the possibility of integrating out all but a finite number of massive spin-2 modes to obtain an effective theory with a strong coupling scale that is parametrically larger than the masses. It translates to an upper bound on the ratio of consecutive eigenvalues of the scalar Laplacian on closed Ricci-flat manifolds with special holonomy, $\lambda_{k+1} \leq 4 \lambda_k$, which applies more generally to closed Einstein manifolds with nonnegative curvature and nonnegative Lichnerowicz spectrum.

We have explored in detail how geometry achieves the nontrivial task of canceling the high-energy growth of massive amplitudes that come from  dimensionally reducing YM theories and GR. Although the amplitudes of single particle states have relatively good high-energy behavior, these theories are still equivalent to higher-dimensional theories and thus are nonrenormalizable. It is natural then to search for theories that are well-behaved deep in the UV. While for spin-1 particles the low-energy dimensionally reduced amplitudes can be reproduced by a quiver theory that admits a standard linear sigma model completion in four dimensions \cite{Hill:2000mu, ArkaniHamed:2001ca, Chivukula:2002ej}, there is no known analogous completion for gravity \cite{ArkaniHamed:2002sp, Scargill:2015wxs}. 
Partial UV completions are known in AdS and from higher-dimensional massive gravity brane constructions~\cite{Porrati:2001db,Porrati:2003sa, Gabadadze:2017jom, Gabadadze:2019lld}. However, a full weakly-coupled UV completion for spin-2 particles seems to require an infinite number of massive higher-spin particles \cite{Camanho:2014apa, Caron-Huot:2016icg,Arkani-Hamed:2017jhn,deRham:2018qqo, Afkhami-Jeddi:2018apj}, as in string theory.

A related question is whether or not any partially unitarized theory of massive spin-2 particles without higher-spin particles must correspond to higher-dimensional gravity with some compact internal geometry. We found that any two-derivative theory with $\sim E^2$ amplitudes and no higher-spin particles must have an infinite number of massive spin-2 particles with GR-like interactions if it has one massive spin-2 particle, so such a theory looks suspiciously like a higher-dimensional gravitational theory.
It would also be interesting if there were ``non-geometric compactifications,'' i.e. solutions to all of the sum rules that do not correspond to the eigenvalues and triple overlap integrals of any manifold.  
Perhaps something like non-geometric string compactifications \cite{Shelton:2005cf,Plauschinn:2018wbo} provide an example.  In our CFT analogy, this would be like CFTs that do not have a Lagrangian description (of which many are known to exist).

Lastly, pure YM and GR amplitudes have myriad interesting properties other than their high-energy behavior, such as the existence of CHY representations, recursion relations, the double copy, and soft theorems, which have been intensely studied, often using dimensional reduction---see Refs.~\cite{Elvang:2013cua, Cheung:2017pzi,Bern:2019prr} for some recent reviews and further references. It would be interesting to further explore how these properties manifest in the lower-dimensional amplitudes of massive KK modes in generic compactifications. 

\paragraph{Acknowledgements:} We would like to thank Clifford Cheung for extensive discussions and early collaboration. We also thank Austin Joyce, Chris Pope and Andrew Tolley for helpful discussions. The authors acknowledge support from DOE grant DE-SC0019143 and Simons Foundation Award Number 658908. We thank the Simons Center for Geometry and Physics for hospitality during the Simons Summer Workshop 2019 during which part of this work was completed.

\appendix

\section{Bounds on eigenvalues}
\label{app:bounds}

In this appendix we briefly review some results about eigenvalues of the scalar Laplace operator, $\Delta$, focusing mainly on closed Riemannian manifolds.  For some reviews of this subject, see Refs.~\cite{chavel1984, LingLu2010}. 

\subsection{Weyl's law}
Weyl's law describes the asymptotic growth of eigenvalues of the scalar Laplacian.
Consider a closed $N$-dimensional manifold $\mathcal{N}$ and let $n(\lambda)$ be the number of eigenvalues of the scalar Laplacian less than $\lambda$ including multiplicities. Weyl's law states that
\be \label{eq:Weyl}
\lim_{\lambda \rightarrow \infty} \frac{n(\lambda) }{\lambda^{N/2}} =  \frac{V}{(4 \pi)^{N/2}\Gamma\left(\frac{N}{2}+1\right)} ,
\ee
where $V$ is the volume of $\mathcal{N}$.
This implies that the spectrum determines the dimension and volume of a manifold.
Weyl's original result was for Dirichlet eigenvalues on bounded domains in $\mathbb{R}^N$ \cite{Weyl1911}.

\subsection{Estimates of the first eigenvalue}
The spectral gap of a manifold is defined as the first nonzero eigenvalue, $\lambda_1$. While the spectral gap can be freely adjusted by rescaling the metric, there are many results bounding it above and below in terms of other geometric quantities. Here we mention a few of these bounds.

Consider a closed manifold $\mathcal{N}$ with dimension $N\geq 2$ such that the Ricci curvature is bounded from below by a positive quantity, i.e. for all vectors $X^m$ we have
\be \label{eq:lichassumption}
(N-1)\kappa g_{mn} X^m X^n \leq R_{mn} X^m X^n,
\ee
where $\kappa >0$. The Lichnerowicz bound then gives a lower bound on the first nonzero eigenvalue \cite{Lichnerowicz},
\be \label{eq:lich1}
N \kappa \leq \lambda_1.
\ee
Obata showed moreover that equality holds in Eq.~\eqref{eq:lich1} only for manifolds that are isometric to the round $N$-sphere \cite{obata1962}. 

We can prove Eq.~\eqref{eq:lich1} by considering an eigenfunction $\psi_a$ with eigenvalue $\lambda_a$ and evaluating
\be
0 \leq \int_{\mathcal{N}} \left( \nabla_n \nabla_m \psi_{a} - \frac{1}{N} \gamma_{nm}\Box \psi_a\right)^2 \leq \frac{(N-1) \lambda_a}{N} \left(\lambda_a-N \kappa \right),
\ee 
where the final inequality comes from integrating by parts and using Eq.~\eqref{eq:lichassumption} together with the fact that $\psi_a$ is an eigenfunction.
For an Einstein manifold, the Lichnerowicz bound gives 
\be \label{eq:lich2}
 \frac{R}{N-1} \leq  \lambda_1.
\ee

Another lower bound for the spectral gap, valid for compact manifolds with nonnegative Ricci curvature, is given by the inequality \cite{LiYau1980, ZhongYang1984}
\be \label{eq:LiYauZhongYang}
\frac{\pi^2}{D^2} \leq \lambda_1,
\ee
where $D$ is the diameter of the manifold, i.e. the supremum of the lengths of the shortest geodesics connecting any two points. This is saturated only by the circle \cite{HangWang2007}, so the inequality is strict for $N\geq 2$. 

There is also an upper bound on the $k$\textsuperscript{th} nonzero eigenvalue $\lambda_k$ for $N$-dimensional compact manifolds with nonnegative Ricci curvature due to Cheng~\cite{Cheng1975}. This says that  
\be
\lambda_k \leq \frac{4 k^2 j^2_{N/2-1,1}}{D^2}<\frac{2k^2 N(N+4)}{D^2},
\ee
where $j_{n,m}$ is the $m$\textsuperscript{th} positive zero of the Bessel function $J_n(x)$. For $k=2$, we can combine this with Eq.~\eqref{eq:LiYauZhongYang} to get a bound on the ratio of the first two nonzero eigenvalues on compact manifolds with nonnegative Ricci curvature,
\be
\frac{\lambda_2}{\lambda_1} \leq \frac{16 j^2_{N/2-1,1}}{\pi^2}.
\ee
The first few values of this upper bound are given approximately in Table~\ref{tab:ratio}.
\begin{table}[ht]
\centering
  \begin{tabular}{ c | c c c c c c }
    $N$ & 1 & 2 & 3& 4 & 5 \\ \hline
     $ 16 \pi^{-2} j^2_{N/2-1,1}$ & 4 & 9.375 & 16 & 23.80 & 32.73
\end{tabular}
\caption{Approximate values of upper bounds on the ratios of the first two nonzero eigenvalues of the scalar Laplacian on $N$-dimensional compact manifolds with nonnegative Ricci curvature.}
\label{tab:ratio}
\end{table}

\subsection{Constraints on all eigenvalues}
\label{sec:deVerdiere}
We now consider some results involving multiple low-lying eigenvalues. 

The first result is due to de Verdi\`ere \cite{deVerdiere1986c, deVerdiere1987a}. It says that given a closed manifold $\mathcal{N}$ of dimension $N \geq 3$ and any finite sequence of non-decreasing positive numbers,
\be \label{eq:seq}
0 <\lambda_1 \leq \lambda_2 \leq \dots \leq \lambda_k,
\ee
then there exists a metric on $\mathcal{N}$ such that \eqref{eq:seq} is the sequence of the first $k$ nonzero eigenvalues of $\Delta$. This shows that the low-lying eigenvalues on a general closed manifold with $N\geq3$ are essentially unconstrained.  Only by making additional assumptions about the manifold can we hope to find non-asymptotic constraints on the eigenvalues, as in the Lichnerowicz bound.  

The next result we mention is the Payne--P\'olya--Weinberger conjecture \cite{PPW56}. The extended conjecture is that the eigenvalues of the scalar Laplacian on a bounded domain $\Omega \subset \mathbb{R}^N$ with Dirichlet boundary conditions satisfy
\be \label{eq:PPW}
\frac{\lambda_{k+1}}{\lambda_k} \bigg|_{\Omega} \leq \frac{\lambda_2}{\lambda_1} \bigg|_{N-{\rm Ball}} = \frac{j^2_{N/2,1}}{j^2_{N/2-1,1}},
\ee
where equality occurs only for the $N$-ball. Approximate values of the upper bound for the first few values of $N$ are given in Table~\ref{tab:PPW}.
\renewcommand\arraystretch{1.5}
\begin{table}[h]
\centering
  \begin{tabular}{ c | c c c c c c }
    $N$ & 1 & 2 & 3& 4 & 5 \\ \hline
     $ \frac{j^2_{N/2,1}}{j^2_{N/2-1,1}}$ & 4 & 2.539 & 2.046& 1.796 & 1.645
\end{tabular}
\caption{Approximate values for the maximum ratios of Dirichlet eigenvalues for some $N$-dimensional balls in Euclidean space.}
\label{tab:PPW}
\end{table}

The case $k=1$ of the conjecture \eqref{eq:PPW} was proven by Ashbaugh and Benguria in Refs.~\cite{ashbaugh1991, ashbaugh1992}. While this result does not directly apply to the compactifications we consider, we review it here as a useful comparison to the eigenvalue bound \eqref{eq:eigenbound} for closed Ricci-flat manifolds with nonnegative Lichnerowicz spectrum.

\renewcommand{\em}{}
\bibliographystyle{utphys}
\addcontentsline{toc}{section}{References}
\bibliography{KK-arxiv-3}

\end{document}